\documentclass[preprint,showpacs,preprintnumbers,amsmath,amssymb,superscriptaddress,10pt,graphicx]{revtex4}

\usepackage{graphicx}
\usepackage{dcolumn}
\usepackage{bm}
\usepackage{amssymb}

\newcommand{\bea}{\begin{eqnarray}}
\newcommand{\ena}{\end{eqnarray}}

\begin{document}

\title{Detection of relic gravitational waves in the CMB: Prospects for CMBPol mission}

\author{Wen Zhao}\email{wzhao7@kasi.re.kr}
\affiliation{International Center for Astrophysics, Korea
Astronomy and Space Science Institute, Daejeon, 305-348,
Korea}\affiliation{School of Physics and Astronomy, Cardiff
University, Cardiff, CF24 3AA, United Kingdom}
\affiliation{Department of Physics, Zhejiang University of
Technology, Hangzhou, 310014, P.R.China}

\date{\today}


\begin{abstract}
{Detection of relic gravitational waves, through their imprint in
the cosmic microwave background radiation, is one of the most
important tasks for the planned CMBPol mission. In the simplest
viable theoretical models the gravitational wave background is
characterized by two parameters, the tensor-to-scalar ratio $r$
and the tensor spectral index $n_t$. In this paper, we analyze the
potential joint constraints on these two parameters, $r$ and
$n_t$, using the potential observations of the CMBPol mission,
which is expected to detect the relic gravitational waves if
$r\gtrsim0.001$. The influence of the contaminations, including
cosmic weak lensing, various foreground emissions, and
systematical errors, is discussed. }

\end{abstract}


\maketitle


\section{Introduction \label{section1}}

The detection of primordial gravitational waves is rightly
considered a highest priority task for the upcoming observation
missions \cite{task}. A stochastic background of the relic
gravitational waves (RGWs), produced in the very early Universe
due to the superadiabatic amplification of zero point quantum
fluctuations of the gravitational field, is a necessity dictate by
general relativity and quantum mechanics \cite{grishchuk}. So the
detection of RGWs maybe provide the unique way to study the birth
of the Universe, and test the applicability of general relativity
and quantum mechanics in the very high energy scale \cite{a12a}.

In a whole range of scenarios of the early Universe, the
primordial power spectrum of the RGWs can be well described by a
power-law form in a fairly large frequency range
\cite{a2a,a2,a3,a4,zhao,others}. Thus, the RGW backgrounds are
conventionally simply characterized by two parameters, the
so-called tensor-to-scalar ratio $r$ and the primordial power
spectral index of RGWs $n_t$, where $r$ describes the amplitude of
the primordial spectrum, and $n_t$ denotes the tilt of the
spectrum. So the constraints on $r$ and $n_t$ will give us a
direct glimpse into the physical conditions in the early Universe.
In particular, they will allow us to place the constraints on the
Hubble parameter of the early Universe, and time evolution of this
Hubble parameter. Unfortunately, the models of the early Universe
cannot give a definite prediction for the values of $r$ and $n_t$,
i.e., different inflationary models predict the quite different
values of $r$ and $n_t$ \cite{liddle}, especially some string
motivated inflationary models predict a very small gravitational
waves with $r\ll 10^{-4}$ \cite{string}. So, the only way to
determine them is by the observations.

The RGWs leave well understood imprints on the anisotropies in
temperature and polarization of cosmic microwave background
radiation (CMB) \cite{polnarev,a8,a10,a11,a13,polnarevTE,zbgte}.
More specifically, RGWs produce a specific pattern of polarization
in the CMB known as the $B$-mode polarization \cite{a8}. Moreover,
RGWs produce a negative cross-correlation between the temperature
and polarization known as the $TE$-correlation at the low
multipoles $\ell\lesssim50$ \cite{polnarevTE,zbgte}. The
theoretical analysis of these imprints along with the data from
CMB experiments allows to place constraints on the parameters $r$
and $n_t$ describing the RGW background, which provides the unique
way to detect the RGWs at the very low frequencies
($10^{-17}\sim10^{-15}$ Hz).

The current CMB experiments are yet to detect a definite signature
of RGWs \cite{currentconstraint}, although a hint of RGWs is found
in the WMAP data \cite{zbghint}. A number of authors have
discussed the possibility of RGW detection by the launched Planck
satellite \cite{Planck,z,zbghint,ma}. The results show that, due
to the fairly large instrumental noises, only if the
tensor-to-scalar ratio is large ($r\gtrsim 0.05$), the Planck
satellite is expected to have a detection. In the previous paper
\cite{zb}, we also found that the Planck mission cannot give a
good constraint for the spectral index $n_t$, even if the
tensor-to-scalar ratio is as large as $r=1$. In addition, various
ground-based \cite{groundbasedCMB} and balloon-borne
\cite{balloonbasedCMB} CMB experiments are expected to have the
better detection abilities, which can constrain the parameter $r$
fairly well when $r\gtrsim0.01$. However, an accurate constraint
of $n_t$ is still unexpected, due to the small partial sky
observation or the short time observation \cite{zz}.

The accurate determination of the RGWs requires the full sky and
the long time observations by the CMB experiment with the quite
small instrumental noises. The future space-based mission focus on
the CMB polarization \cite{cmbpol} (here and in the following, we
use the label `CMBPol' to refer it) provides an excellent
opportunity to realize it (the similar projects, such as B-Pol
\cite{b-pol} and LiteBird \cite{litebird}, are also proposed). The
instrumental noises of CMBPol mission are more than $100$ times
smaller than those of Planck mission. If the foreground
contaminations and the systematical errors can be well controlled,
the signature of RGWs can be well detected, as long as
$r\gtrsim0.001$ \cite{cmbpol}. This will provide an observational
tool to distinguish the different inflationary-type models.

In this paper we shall analyze the joint constraints on two
parameters $r$ and $n_t$ that would be feasible with the analysis
of the observations from the planned CMBPol mission. We shall
detailedly discuss the constraints of $r$, $n_t$ and the
best-pivot wavenumber $k_t^*$ depending on the input (or true)
value of the tensor-to-scalar ratio $r$. We discuss the effects of
various contaminations from the cosmic weak lensing, foreground
radiations and the beam systematics.

The outline of the paper is as follows. In Sec. \ref{section2} we
shall introduce and explain the notations for the power spectra of
gravitational waves, density perturbations and various CMB
anisotropy fields. Furthermore, in this section, we shall briefly
introduce the existence of the best-pivot wavenumber $k_t^*$ for
the detection of RGWs in the CMB. The analytical formulae and
explanation for the $k_t^*$, $\Delta r$ and $\Delta n_t$ will also
be discussed. In Sec. \ref{section3}, by using the analytical
formulae and only considering the instrumental noises, we shall
discuss the values of $k_t^*$, $\Delta r$ and $\Delta n_t$ for
different input (or true) value of $r$. Sec. \ref{section4} is
contributed to show the effect of the cosmic lensing
contamination, and Sec. \ref{section5} is contributed to show the
effect of the foreground radiations contamination. In Sec.
\ref{section6}, we discuss the effects of various beam systematics
for the determination of the parameters $r$ and $n_t$. We also
discuss the requirement of the CMBPol's systematics, if the biases
of the parameters $r$ and $n_t$ are ignorable. Finally, Sec.
\ref{section7} is dedicated to a brief discussion and conclusion.

\section{Optimal parameters and their determinations \label{section2}}

The main contribution to the observed temperature and polarization
anisotropies of the CMB comes from two types of the cosmological
perturbations, density perturbations (also known as the scalar
perturbations) and RGWs (also known as the tensor perturbations)
\cite{polnarev,a3,a4,a8}, which are generally characterized by
their primordial power spectra. These power spectra are usually
assumed to be power-law, which is a generic prediction of a wide
range of scenarios of the early Universe, including the
inflationary models. In general there might be deviations from a
power-law, which can be parameterized in terms of the running of
the spectral index (see for example
\cite{GrishchukSolokhin,liddle}), but we shall not consider this
possibility in the current paper. Thus, the power spectra of the
perturbation fields have the form
 \bea
 P_s(k)=A_s(k_0)(k/k_0)^{n_s-1},~~~~
 P_t(k)=A_t(k_0)(k/k_0)^{n_t},\label{gw-spectrum}
 \ena
for density perturbations and RGWs respectively. In the above
expression $k_0$ is an arbitrarily chosen pivot wavenumber, $n_s$
is the primordial power spectral index for density perturbations,
and $n_t$ is the primordial power spectral index for RGWs.
$A_s(k_0)$ and $A_t(k_0)$ are normalization coefficients
determining the absolute values of the primordial power spectra at
the pivot wavenumber $k_0$. The choices of $n_s=1$ and $n_t=0$
correspond to the scale-invariant power spectra for density
perturbations and gravitational waves respectively.

The relative contribution of density perturbations and
gravitational waves is described by the so-called tensor-to-scalar
ratio $r$, which is defined as follows
\begin{eqnarray} \label{r-define}
r(k_0)\equiv \frac{A_t(k_0)}{A_s(k_0)}.
\end{eqnarray}
Note that, in defining the tensor-to-scalar ratio $r$, we have not
used any inflationary formulae which relate $r$ with the physical
conditions during inflation and the slow-roll parameters (see for
example \cite{stein}). Thus, our definition depends only on the
power spectral amplitudes of density perturbations and RGWs, and
does not assume a particular generating mechanism for these
cosmological perturbations. The RGW amplitude $A_t\left(k_0\right)
= r(k_0)A_s(k_0)$ provides us with direct information on the
Hubble parameter of the very early Universe
\cite{GrishchukSolokhin}. More specifically, this amplitude is
directly related to the value of the Hubble parameter $H$ at a
time when wavelengths corresponding to the wavenumber $k_0$
crossed the horizon \cite{wmap_notation}
\begin{eqnarray}
A_t^{1/2} (k_0)= \left.\frac{\sqrt{2}}{M_{\rm
pl}}\frac{H}{\pi}\right|_{k_0/a = H},\label{rh}
\end{eqnarray}
where $M_{\rm pl}=1/\sqrt{8\pi G}$ is the reduced Planck mass. If
we adopt $A_s=2.445\times 10^{-9}$ as predicted by the WMAP5
observations \cite{wmap5}, the Hubble parameter is
$H\simeq2.67r^{1/2}\times 10^{14}$GeV, only depending on the value
of $r$. In the canonical single-field slow-roll inflationary
models, the Hubble parameter directly relates to the energy scale
of inflation $V^{1/4}$. The relation (\ref{rh}) follows that
$V^{1/4}\simeq3.35r^{1/4}\times10^{16}$GeV, which has been
emphasized by a number of authors.

Assuming that the amplitude of density perturbations $A_s(k_0)$ is
known, taking into account the definitions (\ref{gw-spectrum}) and
(\ref{r-define}), the power spectrum of the RGW field may be
completely characterized by tensor-to-scalar ratio $r$ and the
spectral index $n_t$. It is important to mention that, for
spectral indices different from the scale-invariant case (i.e.,
when $n_s\neq1$ or/and $n_t\neq0$), the definition of the
tensor-to-scalar ratio depends on the pivot wavenumber $k_0$.  If
we adopt a different pivot wavenumber $k_{1}$, the
tensor-to-scalar ratio at this new pivot wavenumber $r(k_1)$ is
related to original ratio $r(k_0)$ through the following relation
(which follows from the definitions (\ref{gw-spectrum}) and
(\ref{r-define}))
\begin{eqnarray}
\label{r-relation} r(k_1)=r(k_0)
\left(\frac{k_1}{k_0}\right)^{n_t-n_s+1}.
\end{eqnarray}

Let us now turn our attention to CMB. Density perturbations and
gravitational waves produce temperature and polarization
anisotropies in the CMB, which are characterized by four angular
power spectra $C_{\ell}^{T}$, $C_{\ell}^{C}$, $C_{\ell}^{E}$ and
$C_{\ell}^{B}$ as functions of the multipole number $\ell$. Here
$C_{\ell}^{T}$ is the power spectrum of the temperature
anisotropies, $C_{\ell}^{E}$ and $C_{\ell}^{B}$ are the power
spectra of the so-called $E$ and $B$ modes of polarization (note
that, density perturbation do not generate $B$-mode of
polarization \cite{a8}), and $C_{\ell}^{C}$  is the power spectrum
of the temperature-polarization cross correlation.

In general, the power spectra $C_{\ell}^{Y}$ (where $Y=T,E,B$ or
$C$) can be presented in the following form
\begin{eqnarray}\label{c-sum}
C_{\ell}^{Y}=C_{\ell}^{Y}({\rm dp})+C_{\ell}^{Y}({\rm gw}),
\end{eqnarray}
where $C_{\ell}^{Y}({\rm dp})$ is the power spectrum due to the
density perturbations, and $C_{\ell}^{Y}({\rm gw})$ is the power
spectrum due to RGWs. In the case of RGWs, the CMB power spectra
can be presented in the following form \cite{a10, a11}
\begin{eqnarray}
C_{\ell}^{Y}({\rm gw})&=&(4\pi)^2 \int \frac{dk}{k} P_{t}(k) \left[\Delta^{(T)}_{Y\ell}(k)\right]^2, ~~{\rm for}~Y=T,E,B, \nonumber\\
C_{\ell}^{C}({\rm gw})&=&(4\pi)^2 \int \frac{dk}{k} P_{t}(k)
\left[\Delta^{(T)}_{T\ell}(k)\Delta^{(T)}_{E\ell}(k)\right].\label{exact-clxx'}
\end{eqnarray}
The transfer functions $\Delta_{Y \ell}^{(T)}(k)$ (see \cite{a10,
a11} for details) in the above expressions translate the power in
the metric fluctuations (gravitational waves) into corresponding
CMB power spectrum at an angular scale characterized by multipole
$\ell$. In this work, for numerical evaluation of the CMB power
spectra due to density perturbations and gravitational waves, we
use the publicly available CAMB code \cite{camb}.

Since we are primarily interested in the parameters of the RGW
field, in the analytical and numerical analysis below we shall
work with a fixed cosmological background model. More
specifically, we shall work in the framework of $\Lambda$CDM
model, and keep the background cosmological parameters fixed at
the values determined by a typical model \cite{wmap5}
\begin{eqnarray}
\label{background} h=0.705,~\Omega_b
h^2=0.02267,~\Omega_{m}h^2=0.1131,~\Omega_{k}=0,~\tau_{reion}=0.084,
~A_s=2.445\times 10^{-9}.
\end{eqnarray}
Furthermore, the spectral indices of density perturbations and
gravitational waves are adopted as follows for the simplicity,
\begin{eqnarray}
n_s=1, ~~n_t=0.
\end{eqnarray}

Note that throughout this paper, we have considered the simplest
cosmological model. In the more general consideration, one should
also include the running of the spectral indices
\cite{GrishchukSolokhin}, the details of the reionization history
\cite{mortonson} and so on, which have been ignored in this paper.

The CMB power spectra $C_{\ell}^{Y}$ are theoretical constructions
determined by ensemble averages over all possible realizations of
the underlying random process. However, in real CMB observations,
we only have access to a single sky, and hence to a single
realization. In order to obtain information on the power spectra
from a single realization, it is required to construct estimators
of power spectra. In order to differentiate the estimators from
the actual power spectra, we shall use the notation $D_{\ell}^Y$
to denote the estimators while retaining the notation $C_{\ell}^Y$
to denote the power spectrum. It is important to keep in mind that
the estimators $D_{\ell}^Y$ are constructed from observational
data, while the power spectra $C_{\ell}^Y$ are theoretically
predicted quantities. The probability distribution functions for
the estimators are described in detail in \cite{zbgte} (see also
\cite{z,wishart2,pdf1}), which predicts the expectation values of
the estimators
\begin{equation} \langle D_{\ell}^Y\rangle=C_{\ell}^Y,\label{mean}
\end{equation}
and the standard deviations
\begin{eqnarray}
(\sigma_{D_{\ell}^X})^2&=&\frac{2(C_{\ell}^X+N_{\ell}^X)^2}{(2\ell+1)f_{\rm sky}},~~(X=T,E,B)\nonumber\\
(\sigma_{D_{\ell}^C})^2&=&\frac{(C_{\ell}^T+N_{\ell}^T)(C_{\ell}^E+N_{\ell}^E)+(C_{\ell}^C+N_{\ell}^C)^2}{(2\ell+1)f_{\rm
sky}},\label{variance}
\end{eqnarray}
where $f_{\rm sky}$ is the sky-cut factor. In this paper, we use
$f_{\rm sky}=0.8$ for the CMBPol survey. $N_{\ell}^Y$ are the
noise power spectra, which are all determined by the specific
experiments. In this formulas, the possible bias generated by the
beam systematics has not been considered (see Sec. \ref{section6}
for details).

In order to estimate the parameters $r$ and $n_t$ characterizing
the RGW background, we shall use an analysis based on the
likelihood function \cite{cosmomc}. The likelihood function is
just the probability density function of the observational data
considered as a function of the unknown parameters (which are $r$
and $n_t$ in our case). Up to a constant, independent of its
arguments, the likelihood function is given by
\begin{eqnarray}\label{likelihood}
\nonumber \mathcal{L} = \prod_{\ell}
f(D_{\ell}^{C},D_{\ell}^{T},D_{\ell}^{E},D_{\ell}^{B}),
\end{eqnarray}
where the function
$f(D_{\ell}^{C},D_{\ell}^{T},D_{\ell}^{E},D_{\ell}^{B})$ is
explained in detail in the previous works \cite{z,zb}.

In the previous works \cite{z,zb}, we have discussed how to
constrain the parameters of the RGWs, $r$ and $n_t$, by the CMB
observation. In \cite{zb}, we found that in general, the
constraints on $r$ and $n_t$ correlate with each other. However,
if we consider the tensor-to-scalar ratio at the best-pivot
wavenumber $k_t^*$, i.e. $r^*\equiv r(k_t^*)$, the constraints on
$r$ and $n_t$ becomes independent of each other, and the
uncertainties $\Delta{r}$ and $\Delta{n_t}$ have the minimum
values. We have derived the formulae to calculate the quantities:
the best-pivot wavenumber $k_t^*$, and the uncertainties of the
parameters $\Delta r$ and $\Delta {n_t}$, which provides a simple
and quick method to investigate the detection abilities of the
future CMB observations. We shall briefly introduce these results
in this section.

It is convenient to define the quantities as below,
 \bea
 a_{\ell}^{Y}\equiv \frac{C_{\ell}^Y({\rm gw})}{\sigma_{D_{\ell}^Y}},~~b^*_{\ell}\equiv
 \ln\left(\frac{\ell}{\ell_t^*}\right),~~d_{\ell}^Y\equiv
 \frac{D_{\ell}^Y-C_{\ell}^Y({\rm dp})}{\sigma_{D_{\ell}^Y}},
 \ena
where $C_{\ell}^Y({\rm gw})$ is the CMB power spectrum generated
by RGWs, and $\sigma_{D_{\ell}^Y}$ is the standard deviation of
the estimator $D_{\ell}^Y$, which can be calculated by Eq.
(\ref{variance}). We should notice that, the quantity $d_{\ell}^Y$
is dependent of random date $D_{\ell}^Y$. By considering the
relations in (\ref{mean}) and (\ref{c-sum}), we can obtain that
$\langle d_{\ell}^Y\rangle=a_{\ell}^Y$, which shows that
$d_{\ell}^Y$ is an unbiased estimator of $a_{\ell}^Y$. $\ell_t^*$
is the so-called best-pivot multipole, which is determined by
solving the following equation \cite{zb}:
 \bea\label{ltstar}
 \sum_{\ell}\sum_{Y}a_{\ell}^{Y2}b_{\ell}^*=0.
 \ena
So the value of $\ell_t^*$ depends on the cosmological model, the
amplitude of RGWs, and noise power spectra by the quantity
$a_{\ell}^Y$. The best-pivot wavenumber $k_t^*$ relates to
$\ell_t^*$ by the approximation relation \cite{zb},
\begin{equation}\label{kl}
k_t^*\simeq \ell_t^*\times 10^{-4}{\rm Mpc}^{-1}.
\end{equation}
The numerical factor here mainly reflects the angular-diameter
distance to the last scattering surface.

\begin{figure}
\begin{center}
\includegraphics[width=12cm,height=10cm]{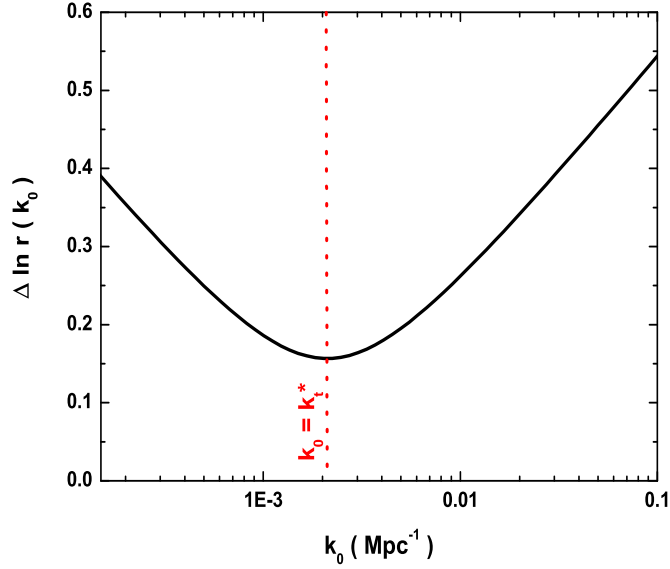}
\end{center}\caption{ The uncertainty of $r(k_0)$ for the different pivot wavenumber $k_0$.}\label{figure0}
\end{figure}

Once the value of $\ell_t^*$ is obtained, the uncertainties
$\Delta r^*$ and $\Delta n_t$ can be calculated by the following
simple formulae
 \bea\label{uncertainties}
 \Delta r^*=r^*/\sqrt{\sum_{\ell}\sum_{Y}a_{\ell}^{Y2}},~~~~
 \Delta
 n_t=1/\sqrt{\sum_{\ell}\sum_{Y}(a_{\ell}^{Y}b_{\ell}^*)^2}.
 \ena
As usual, we can define the signal-to-noise ratio $S/N\equiv
r^*/\Delta r^*$. Using (\ref{uncertainties}), we get
 \bea\label{snr}
 S/N=\sqrt{\sum_{\ell}\sum_{Y}a_{\ell}^{Y2}}.
 \ena
In the previous work \cite{zb}, we found the uncertainty of
$r(k_0)$, the tensor-to-scalar ratio at the pivot wavenumber
$k_0\neq k_t^*$, is larger than $\Delta r^*$. The value of $\Delta
\ln r(k_0)$ is fairly well approximated by the following formula
 \bea
 \Delta \ln r(k_0)=\sqrt{(\Delta r^*/r^*)^2+(\ln(k_0/k_t^*)\Delta
 n_t)^2}.
 \ena
The smallest uncertainty on tensor-to-scalar ratio $r$ is achieved
for the choice of the pivot scale at $k_0=k_t^*$. This justified
the title `best' pivot wavenumber for $k_t^*$. We should notice
that the values of $k_t^*$, $S/N$, $\Delta \ln r$ and $\Delta n_t$
only depend on the input (or true) cosmological model, but not on
the data $D_{\ell}^Y$. In Fig. \ref{figure0}, we plot the value of
$\Delta \ln r(k_0)$ as a function of the pivot scale $k_0$, where
the input model has $r=0.3$, and the Planck instrumental noises
are considered (see \cite{zb} for details). As expected, when
$k_0\gg k_t^*$ or $k_0\ll k_t^*$, the uncertainty becomes much
larger than $\Delta r^*$.

The likelihood function in (\ref{likelihood}) has the maximum
value at $(r^*_{\rm ML}, n_{t{\rm ML}})$. The values of $r^*_{\rm
ML}$ and $n_{t{\rm ML}}$ depend on the data $D_{\ell}^{Y}$,
different from the value of $\Delta r^*$ and $\Delta n_t$. In the
previous work \cite{zb}, we found that the values of $r^*_{\rm
ML}$ and $n_{t{\rm ML}}$ can be very well approximated by the
follows
 \bea
  r^*_{\rm ML}=r^*\frac{\sum_{\ell}\sum_{Y}a_{\ell}^Yd_{\ell}^Y}{\sum_{\ell}\sum_{Y}a_{\ell}^{Y2}},
~~~
 n_{t{\rm
 ML}}=n_t+\frac{\sum_{\ell}\sum_{Y}a_{\ell}^Yd_{\ell}^Yb_{\ell}^*}{\sum_{\ell}\sum_{Y}(a_{\ell}^{Y}b_{\ell}^*)^2},
 \ena
which depend on the data by the quantity $d_{\ell}^Y$. If the CMB
estimator $D_{\ell}^{Y}$ is unbiased for $C_{\ell}^{Y}$, as
discussed above, we have $\langle r^*_{\rm ML}\rangle=r^*$ and
$\langle n_{t{\rm ML}}\rangle=n_t$, where Eq. (\ref{ltstar}) is
used. These show that $r^*_{\rm ML}$ and $n_{t{\rm ML}}$ are the
unbiased estimators for $r^*$ and $n_t$, respectively. However,
when $D_{\ell}^{Y}$ is a biased estimator for $C_{\ell}^{Y}$,
$r^*_{\rm ML}$ and $n_{t{\rm ML}}$ will also be the biased
estimators for $r^*$ and $n_t$, respectively (see Sec.
\ref{section6} for details), which brings the errors for the
detection of RGWs.

The detection ability of the CMB experiment strongly depends on
the noise levels, which include the instrumental noises, cosmic
lensing contaminations, foreground radiation contaminations and
the beam systematics. In the following sections, we shall discuss
these effects separately. In addition, due to the partial sky
survey, the leakage from the $E$-polarization into the
$B$-polarization could be another kind of contamination. However,
it was found that, this $E$-$B$ mixture can be properly avoided
(or deeply reduced) by constructing the pure $E$-mode and $B$-mode
polarization fields (see \cite{cutsky1,cutsky2,cutsky3} for
details). So we shall not discuss this topic in this paper.

\section{CMBPol instrumental noises' contaminations \label{section3}}

\begin{figure}
\begin{center}
\includegraphics[width=12cm,height=10cm]{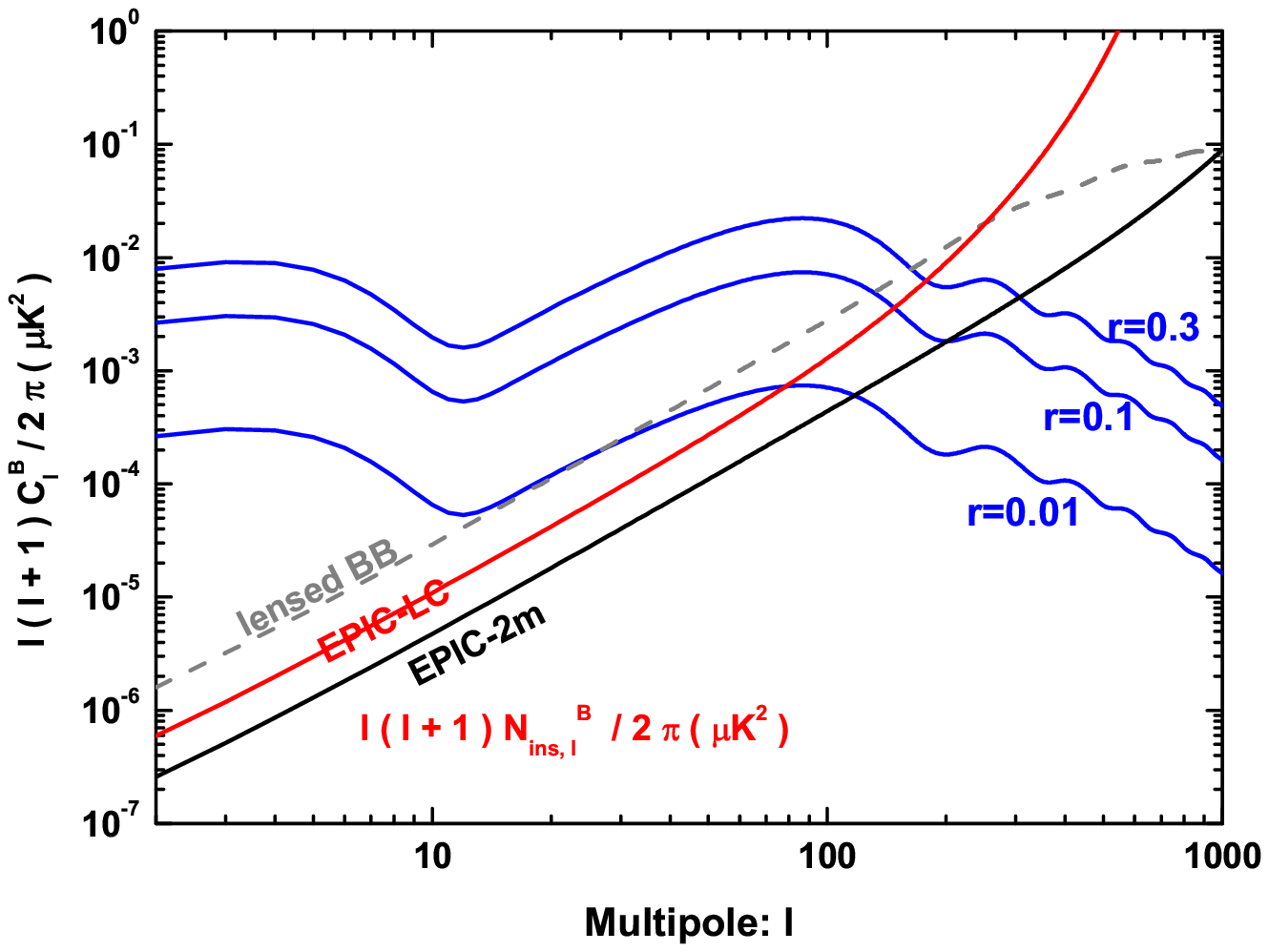}
\end{center}\caption{The instrumental noise power spectra $N_{{\rm ins},\ell}^{B}$ of EPIC-2m (black line) and EPIC-LC (red line).}\label{figure1}
\end{figure}

\begin{figure}
\begin{center}
\includegraphics[width=18cm,height=10cm]{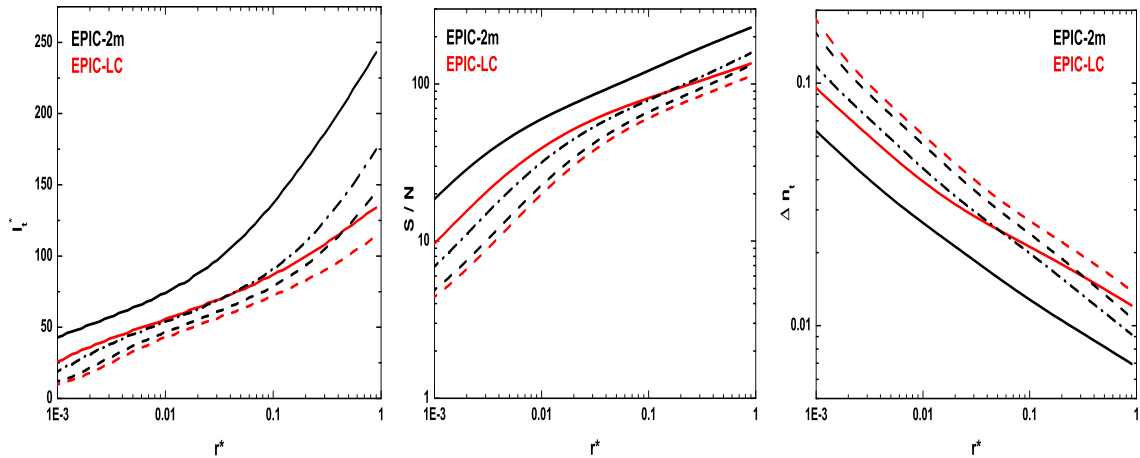}
\end{center}\caption{ The figures show the values of the best-pivot multipole $\ell_t^*$ (left panel),
signal-to-noise ratio $S/N$ (middle panel) and the uncertainty of
the RGW spectral index $\Delta n_t$ (right panel) as functions of
$r^*$. The black solid (dash-dotted, dashed) lines correspond to
the EPIC-2m instrumental noises (instrumental noises $+$ the
reduced cosmic lensing contaminations, instrumental noises $+$
cosmic lensing contaminations), and the red solid (dashed) lines
correspond to the EPIC-LC instrumental noises (instrumental noises
$+$ the cosmic lensing contaminations).}\label{figure2}
\end{figure}

\begin{figure}
\begin{center}
\includegraphics[width=16cm,height=10cm]{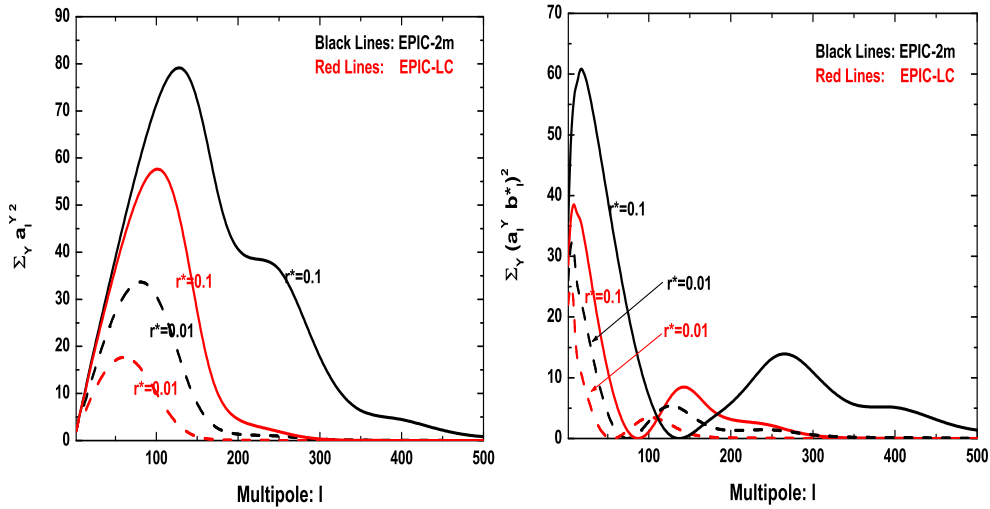}
\end{center}\caption{The figures show the values of $\sum_{Y}a_{\ell}^{Y2}$ (left panel) and
$\sum_{Y}(a_{\ell}^{Y}b^*_{\ell})^2$ (right panel) as functions of
multipole $\ell$ for the cases with the different $r^*$. Here, we
have not considered the contaminations from cosmic weak lensing
and foreground emissions. }\label{figure3}
\end{figure}

\begin{table}
\caption{Experimental specifications for the mid-cost
 (EPIC-2m) CMBPol mission and the low-cost
 (EPIC-LC) CMBPol mission. \cite{cmbpol} }
\begin{center}
\label{table1}
\begin{tabular}{|c|c|c|c|c|c|c|}
    \hline
    & Frequency~[GHz] & ~~45~~ & ~~70~~ & ~~100~~ & ~150~ & ~220~  \\
    EPIC-2m& ${\rm \theta}_{\rm F}$~[arcmin] & ~~17~~ & ~~11~~ & ~~8~~ & ~5~ & ~3.5~  \\
   & $\Delta_T$~[${\rm \mu K}$-${\rm arcmin}$] & ~~5.85~~ & ~~2.96~~ & ~~2.29~~ & ~2.21~ & ~3.39~  \\
  \hline
    & Frequency~[GHz] & ~~40~~ & ~~60~~ & ~~90~~ & ~135~ & ~200~  \\
   EPIC-LC   & ${\rm \theta}_{\rm F}$~[arcmin] & ~~116~~ & ~~77~~ & ~~52~~ & ~34~ & ~23~  \\
   & $\Delta_T$~[${\rm \mu K}$-${\rm arcmin}$] & ~~15.27~~ & ~~8.23~~ & ~~3.56~~ & ~3.31~ & ~3.48~  \\
  \hline
\end{tabular}
\end{center}
\end{table}

In this section, we shall discuss the determination of RGWs, when
only taking into account the instrumental noises of the CMBPol
mission.

For a single frequency channel $i$, we assume Gaussian beams. The
noise power spectrum (after deconvolution of the beam window
function) is
 \bea
 N_{{\rm ins},\ell}^{T}(i)=(\Delta_T)^2\exp\left[\frac{\ell(\ell+1)\theta_F^2}{8\ln
 2}\right],~~N_{{\rm ins},\ell}^{C}(i)=0,
 \ena
and
 \bea
 N_{{\rm ins},\ell}^{E}(i)=N_{{\rm ins},\ell}^{B}(i)=(\Delta_P)^2\exp\left[\frac{\ell(\ell+1)\theta_F^2}{8\ln
 2}\right],
 \ena
where $\theta_{\rm F}$ is the full width at half maximum (FWHM) of
the beam $i$. $\Delta_T$ and $\Delta_P$ are the noises for the
temperature and polarizations, which relate by
$\Delta_P=\sqrt{2}\Delta_T$. The values of the $\Delta_T$ and
$\Delta_P$ depend on the number of the detectors, the integration
time and the survey area.

If the experiment includes several different channels, we need to
generalize the above considerations. The optimal channel
combination of these channels gives the total effective
instrumental noise \cite{cmbpol},
 \bea
 [N_{{\rm ins},\ell}^{Y}]^{-1}=\sum_{i}\left[N_{{\rm ins},\ell}^{Y}(i)\right]^{-1},
 \ena
where $i$ runs though the channels, $N_{{\rm ins},\ell}^{Y}(i)$ is
the instrumental noise bias of the channel $\nu_i$. In this
section, we shall only consider the instrumental noises, i.e.
 \bea\label{ins-noise}
 N_{\ell}^{Y}\mapsto N^{Y}_{{\rm ins},\ell}.
 \ena

Since the precise experimental specifications of CMBPol have not
yet been defined, we will consider two different cases (EPIC-2m)
and (EPIC-LC) suggested by \cite{cmbpol} (recently, an
EPIC-Intemediate Mission is also suggested by the CMBPol team
\cite{cmbpol-middle}). The experimental specifications are given
in Table \ref{table1}, where 2-year design life is assumed
\footnote{Note that in the real analysis throughout this paper,
similar to \cite{cmbpol}, we have not considered the frequency
channels at $30$GHz and $340$GHz proposed in \cite{cmbpol} for
EPIC-2m, and the frequency channels at $30$GHz and $300$GHz for
EPIC-LC.}. In Fig. \ref{figure1}, we plot the polarization noise
spectra $N_{{\rm ins},\ell}^{B}$ of EPIC-2m and EPIC-LC,
respectively \footnote{Note that, in Sec. \ref{section5}, we will
show that the total noise level will be increased if the reduced
foreground contaminations are considered. However, the noise level
with high multipole $\ell>100$ nearly keeps same.}. For EPIC-2m,
when $\ell<100$, we find that $N_{{\rm ins},\ell}^B\sim
2.7\times10^{-7}\mu$K$^2$, which is nearly $400$ times smaller
than that of the Planck mission (7 frequency channels from $30$GHz
to $353$GHz and $28$-month surveying time are assumed
\cite{Planck}). Even for the EPIC-LC, when $\ell<100$, we have
$N_{{\rm ins},\ell}^B\sim 6.2\times10^{-7}\mu$K$^2$, $200$ times
smaller than that of the Planck mission. So comparing with Planck
mission, CMBPol is much more sensitive for the detection of CMB
polarization. From Fig. \ref{figure1}, we also find that even for
the model with quite small $r=0.01$, the value of $N_{{\rm
ins},\ell}^B$ is smaller than that of $C_{\ell}^{B}$ when
$\ell<120$ for EPIC-2m, and when $\ell<80$ for EPIC-LC. So the
CMBPol mission provides an excellent opportunity to detailedly
observe the peak of $C_{\ell}^{B}$ at $\ell\sim80$.

Let us discuss the constraints on the gravitational waves by the
potential observations of CMBPol mission. We shall discuss the
values of the best-pivot scale $k_t^*$, the signal-to-noise ratio
$S/N$ and the uncertainty of the spectral index $\Delta n_t$, by
considering the CMBPol instrumental noises.

The value of $k_t^*$ directly relates to the best-pivot multipole
$\ell_t^*$ by Eq.(\ref{kl}), and the value of $\ell_t^*$ is
obtained by solving the equation in (\ref{ltstar}). By using
(\ref{ins-noise}), we obtain the value of $\ell_t^*$ as a function
of the input (or true) value of the tensor-to-scalar ratio $r^*$
for EPIC-2m and EPIC-LC, which are plotted in Fig. \ref{figure2}
(left panel). We find that, in both cases, the value of $\ell_t^*$
becomes larger with the increasing of $r^*$. For EPIC-2m, we have
$\ell_t^*=43$ for $r^*=0.001$, and $\ell^*_t=137$ for $r^*=0.1$.
For EPIC-LC, the value of $\ell_t^*$ is smaller than that of
EPIC-2m, due to the larger noise level and the larger beam FWHM of
EPIC-LC. When $r=0.001$, we have $\ell_t^*=26$, and when $r=0.1$,
we have $\ell_t^*=87$. These reflect that gravitational waves in
the frequency range $k\sim0.01$Mpc$^{-1}$ will be best constrained
by the future CMBPol observations, unless the value of $r$ is
extremely small. This is because the main contribution comes from
the observation of the peak of $B$-polarization at $\ell\sim80$.
We should remember that this is different from the Planck case,
where $\ell_t^*\sim 10$, due to the main contribution of the
reionization peak of $B$-polarization \cite{zbgte,z,zb}.

The signal-to-noise ratio is calculated by Eq. (\ref{snr}). By
using Eq. (\ref{ins-noise}), we get the value of $S/N$ as a
function of $r^*$ for both EPIC-2m and EPIC-LC, which are shown in
Fig. \ref{figure2} (middle panel). As expected, the signal of RGWs
can be very well determined by the CMBPol mission. Even for the
model with $r=0.001$, we can have $S/N=19$ for EPIC-2m and
$S/N=10$ for EPIC-LC, when only considering the corresponding
instrumental noises. When $r=0.1$, we have $S/N=122$ for EPIC-2m
and $S/N=81$ for EPIC-LC.

We can also calculate the value of $\Delta n_t$ by using Eq.
(\ref{uncertainties}) and the value of $\ell_t^*$ given in left
panel of Fig. \ref{figure2}. The results are shown in Fig.
\ref{figure2} (right panel). As expected, the value of $\Delta
n_t$ decreases with the increasing of $r^*$. For EPIC-2m, we have
$\Delta n_t=0.06$ for the model with $r^*=0.001$, and $\Delta
n_t=0.01$ for the model with $r^*=0.1$. This uncertainty is about
$20$ times smaller than that given by Planck satellite
\cite{zb,zz}. This constraint, combining with $\Delta r^*$, will
give a quite sensitive way to differentiate various inflationary
type models. For the EPIC-LC, the uncertainty of $n_t$ is about
$2$ times larger that of EPIC-2m. When $r=0.001$, we have $\Delta
n_t=0.10$, and when $r=0.1$, we have $\Delta n_t=0.02$.

It is necessary to discuss the contributions of $S/N$ and $\Delta
n_t$ from every multipole, which can be very easily analyzed by
the analytical formulae. From Eqs. (\ref{uncertainties}) and
(\ref{snr}), we find these two quantities can be rewritten as
follows
 \bea\label{fenjie}
 (S/N)^2=\sum_{\ell}\sum_{Y}a_{\ell}^{Y2},~~~(1/\Delta
 n_t)^2=\sum_{\ell}\sum_{Y}(a_{\ell}^{Y}b_{\ell}^*)^2.
 \ena
They are the simple sums of the contributions from each multipole
$\ell$ and CMB information channel $Y$. We plot the functions of
$\sum_{Y}a_{\ell}^{Y2}$ and $\sum_{Y}(a_{\ell}^{Y}b_{\ell}^*)^2$
as a function of $\ell$ for two different models ($r=0.01$ and
$r=0.1$). The results are shown in Fig. \ref{figure3}. Left panel
shows that all these four lines are peaked at $\ell\sim100$, which
is close to the peak of $B$-polarization. This reflects that when
$r>0.01$, the main contribution comes from the observation in the
range $\ell\sim100$, consistent with our previous discussion. This
is different from the case of Planck satellite \cite{zbgte,z},
where the reionization peak at $\ell\sim 6$ is extremely
important. In the CMBPol case, the contribution from the largest
scale $\ell<20$ is unimportant due to the cosmic variance, and the
contribution from the small scale $\ell>300$ is also unimportant
for the large instrumental noises. However, it is important to
mention that if $r\ll0.01$, similar to Planck satellite, the
reionization peak at $\ell<10$ again becomes the main contribution
for the total $S/N$.

However, it is different for $\sum_{Y}(a_{\ell}^{Y}b_{\ell}^*)^2$,
which stands for the individual contribution for $1/\Delta n_t$.
From the right panel of Fig. \ref{figure3}, we find that this
function is sharply peaked at the largest scale $\ell<30$, and the
contribution from intermedial scale around the best-pivot multpole
is very small. These can be easily understood, the quantity
$b_{\ell}^*\equiv \ln(\ell/\ell_t^*)$ is zero when
$\ell=\ell_t^*$, which follows that
$\sum_{Y}(a_{\ell}^{Y}b_{\ell}^*)^2=0$ at $\ell=\ell_t^*$. Only if
$\ell\ll\ell_t^*$ or $\ell\gg\ell_t^*$, $b_{\ell}^*$ has a large
value, and follows a large $\sum_{Y}(a_{\ell}^{Y}b_{\ell}^*)^2$.
Especially, the contribution from $\ell\ll\ell_t^*$ is very
important. For example, when $\ell_t^*=137$, $b^{*2}_{\ell=2}$ is
$30$ times large than $b^{*2}_{\ell=300}$. This reflects that the
constraint on the tilt of the primordial gravitational waves power
spectrum strongly depends on the observations in a large scale
range. The cosmic reionization is very important for the
constraint of $n_t$, although it might not be so important for the
constraint of $r$ for the CMBPol observations.

\section{Cosmic weak lensing contamination \label{section4}}

\begin{figure}
\begin{center}
\includegraphics[width=16cm,height=10cm]{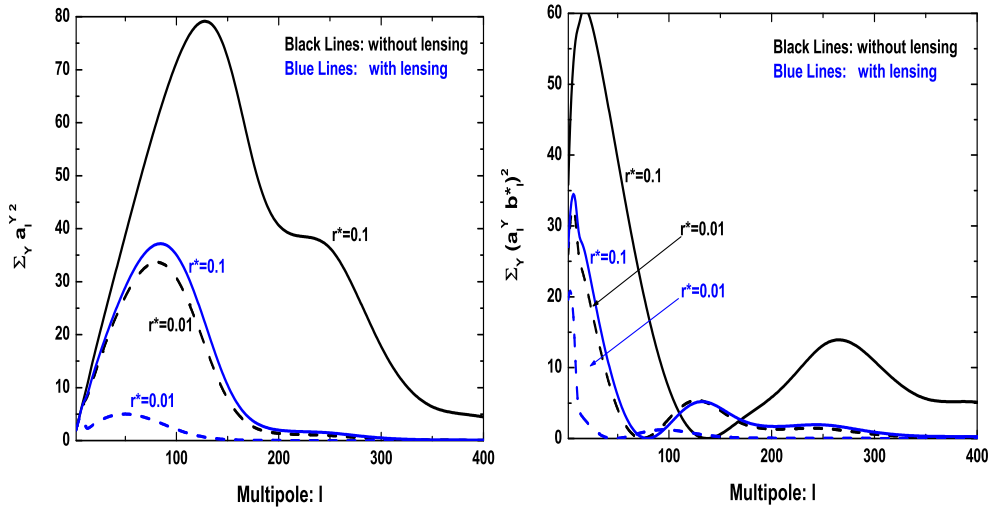}
\end{center}\caption{For EPIC-2m, the figures show the values of $\sum_{Y}a_{\ell}^{Y2}$ (left panel) and
$\sum_{Y}(a_{\ell}^{Y}b^*_{\ell})^2$ (right panel) as functions of
multipole $\ell$ for the cases with and without the reduced cosmic
lensing contamination with the residual factor $\sigma^{\rm
lens}=0.5$. }\label{figure31}
\end{figure}

In \cite{a8}, it was pointed out that the gravitational waves
result in CMB polarization with a $B$-mode, whereas density
perturbations do not. Thus, the signal of gravitational waves
could not be confused with density perturbations by detecting the
$B$-polarization. Although, the amplitude of the $B$-polarization
is expected to be quite small, it gives a clear information for
gravitational waves. However, when taking into account the
second-order effect, the $B$-mode can also arise from the lensing
of the $E$-mode by density perturbations along lines-of-sight
between the observer and the last-scattering surface
\cite{lensing0}. The scalar contribution to the $B$-mode power
spectrum is shown in Fig. \ref{figure1} (grey dashed line). When
$\ell<200$, it is nearly a white spectrum with the amplitude
$C_{{\rm lens},\ell}^B\simeq 2\times 10^{-6}\mu$K$^2$, which is
$7$ times larger than the instrumental noises of EPIC-2m, and $3$
times larger than that of EPIC-LC.

When the instrumental noise of the CMB experiment is sufficiently
small, as the CMBPol mission, the gravitational lensing
contribution to the large-scale $B$-mode becomes one of the
limiting sources of contamination for constraining the RGWs.
High-sensitivity measurements of small-scale $B$-modes can reduce
this contamination through a lens reconstruction technique, which
has been discussed by a number of authors (see for instance
\cite{lensing1,lensing2,wuran4}).

The effect of cosmic lensing contamination for the detection of
RGWs can be easily discussed. The reduced lensed $B$-mode
polarization can be treated simply as a well-known noise for
gravitational waves in the likelihood analysis, i.e.
 \bea
 N_{\ell}^{B} \mapsto N_{{\rm
 ins},\ell}^{B}+C_{{\rm lens},\ell}^{B}\times \sigma^{\rm lens},
 \ena
where we have defined the residual factor $\sigma^{\rm lens}$ for
the lensed $B$-polarization. Note that in the real situation, the
lens-induced $B$-modes are non-Gaussian, so we should not behave
exactly it as the additional Gaussian noise (see \cite{wuran4} and
references therein). However, on the scales relevant for $B$-mode
detection in this paper, the non-Gaussianity has only a minor
effect, which has been ignored in our discussion. In general, we
have $\sigma^{\rm lens}\le1$, with the equality holding for the
lensed $B$-mode is not reduced. The reduction of gravitational
lensing contribution strongly depends on the instrumental noise
level, the beam FWHM, the foregrounds and the instrumental
systematics. In the work \cite{cmbpol-lensing}, the authors found
that, based on the noise level of EPIC-2m, one can expect to have
$\sigma^{\rm lens}\sim0.5$. However, for EPIC-LC, it is very
difficult to reduce the cosmic lensing contamination due to the
large beam FWHM. We should mention that since the value of
$C_{{\rm lens},\ell}^{B}$ is much larger than the instrumental
noises of CMBPol mission, in the total effective noise
$N_{\ell}^{B}$, the cosmic lensing contamination becomes the
dominant portion.

We have calculated the constraints of the gravitational waves by
taking into account the cosmic lensing contaminations. The values
of $\ell_t^*$, $S/N$ and $\Delta n_t$ are shown in Fig.
\ref{figure2}, where $\sigma^{\rm lens}=0.5$ and $\sigma^{\rm
lens}= 1$ are considered for EPIC-2m, and $\sigma^{\rm lens}= 1$
is considered for EPIC-LC. The left panel shows that, the
best-pivot multipole is shifted to smaller scale by the lensing
contamination. We find the value of $S/N$ is much reduced by the
lensing contamination, especially for the case with small
tensor-to-scalar ratio. When $r=0.001$, we have $S/N=7$ for
EPIC-2m (with $\sigma^{\rm lens}= 0.5$) and $S/N=4$ for EPIC-LC,
which are much smaller than the corresponding values with only
instrumental noises. The uncertainty of $n_t$ is also much
increased by the cosmic lensing, especially for the case with
small $r$. When $r=0.001$, EPIC-2m with $\sigma^{\rm lens}= 0.5$
can give $\Delta n_t=0.12$, and EPIC-2m with $\sigma^{\rm lens}=
1$ can give $\Delta n_t=0.16$, which are more than $2$ times
larger than those in the case without cosmic lensing, and are
fairly loose to differentiate various inflationary models.
However, when the tensor-to-scalar ratio is $r=0.1$, we have
$\Delta n_t=0.02$ for EPIC-2m, which is still a quite tight
constraint.

We have also investigated the contribution of every multipole for
$S/N$ and $\Delta n_t$. The results can be found in Fig.
\ref{figure31}, where we have focused on the EPIC-2m mission and
$\sigma^{\rm lens}=0.5$ is used for the case with cosmic lensing
contamination. We find that the peak in each case is much reduced
by the lensing contamination.

It is interesting to mention that for the high-sensitivity
detectors the residual lensing noise dominates over the
instrumental noises, and place the detection limit for CMB
experiments \cite{song}. In \cite{lensing2}, the authors claimed
that, for a extreme high-sensitivity detector, a reduction in
lensing power by a factor $40$ is possible using approximate
iterative maximum-likelihood method. If we consider this residual
as the lower limit of the reduced lensing noises, we find that
$r>3.7\times 10^{-6}$ can be detected at more than $2$-$\sigma$
level in absence of sky cuts, foregrounds and instrumental
systematics \cite{zb}. This can be treated as the detection limit
of the CMB experiments for gravitational waves. This lower limit
corresponds to the Hubble parameter $H\simeq 3.1\times10^{11}$GeV,
and the energy scale of inflation
$V^{1/4}\simeq1.5\times10^{15}$GeV. In this limit case, the
uncertainty of spectral index also becomes very small. When
$r=0.1$, we have $\Delta n_t=0.007$ (see Fig. 2 in \cite{zb} for
details), placing a very tight constraint on the inflationary
models.

\section{Foreground contaminations \label{section5}}

\begin{table}
\caption{Assumptions about foreground emissions. \cite{cmbpol} }
\begin{center}
\label{table2}
\begin{tabular}{|c|c|c|}
    \hline
    Parameter &~~~{\bf Synchrotron}~~~&~~~{\bf Dust}~~~   \\
    \hline
    $A_{S,D}$  & $4.7\times 10^{-5}$~$\mu$K$^{2}$ &  $1.2\times10^{-4}$~$\mu$K$^{2}$   \\
    \hline
    $\nu_0$   & 30~GHz & 94~GHz   \\
    \hline
    $\ell_0$   & 350 & 900   \\
    \hline
    $\alpha$   & -3 & 2.2   \\
    \hline
    $\beta^{E}$   & -2.6 & -1.3   \\
    \hline
    $\beta^{B}$   & -2.6 & -1.4   \\
    \hline
    $\beta^{C}$   & -2.6 & -1.95   \\
  \hline
\end{tabular}
\end{center}
\end{table}

\begin{figure}
\begin{center}
\includegraphics[width=8cm,height=8cm]{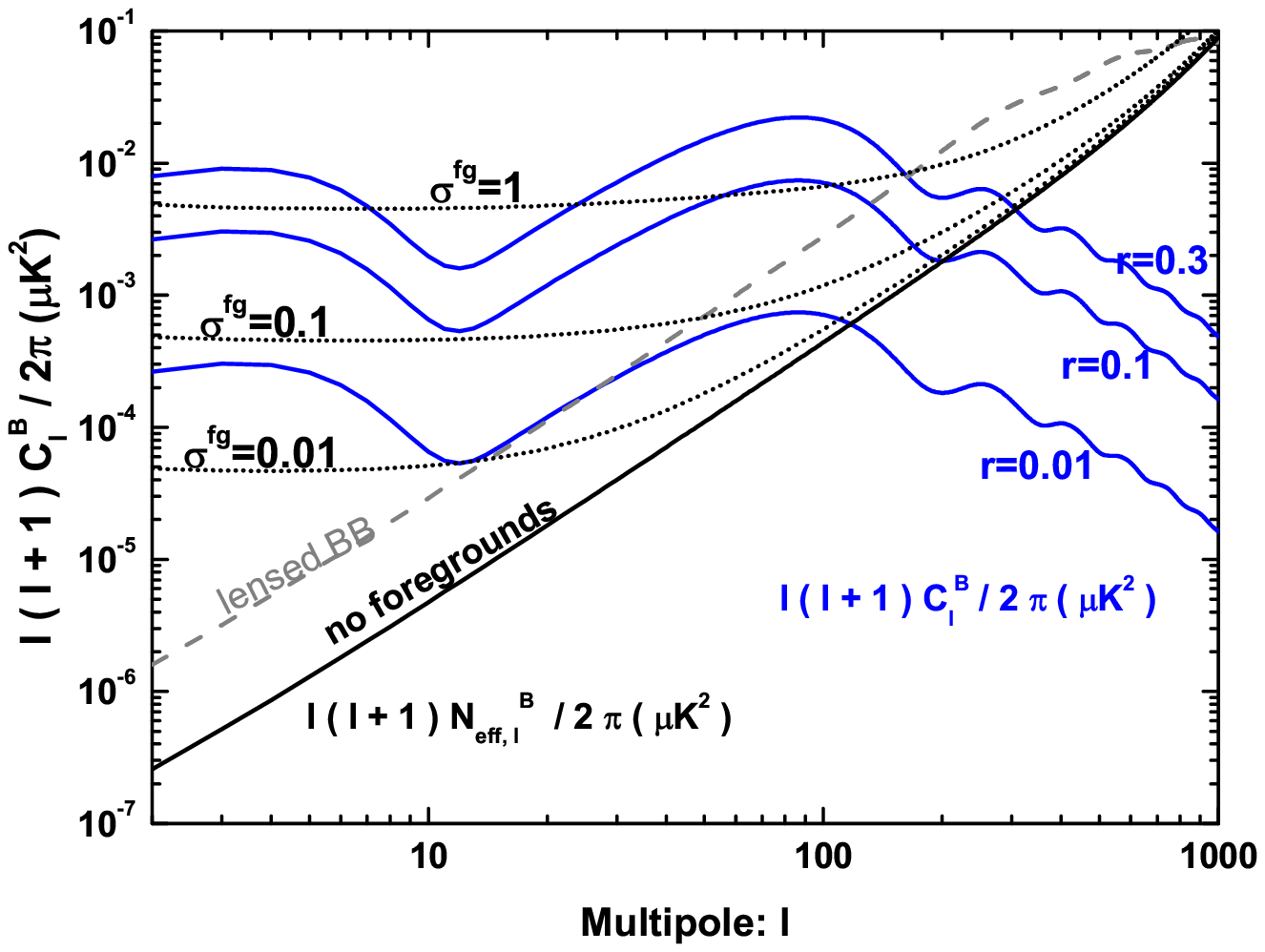}\includegraphics[width=8cm,height=8cm]{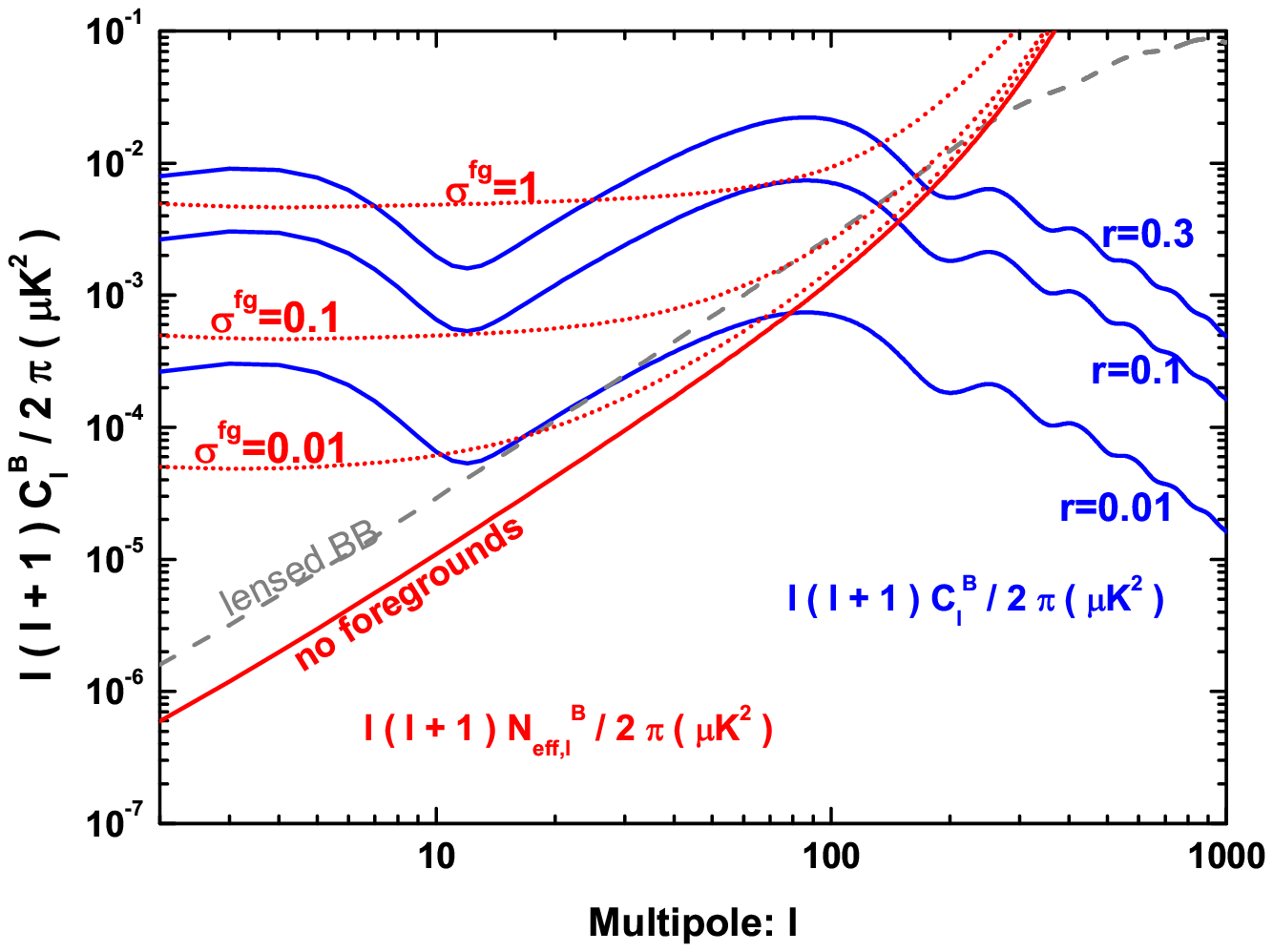}
\end{center}\caption{The total effective noise power spectra $N_{{\rm eff},\ell}^{B}$ when considering the foreground
contaminations. The left panel is for EPIC-2m and the right panel
is for EPIC-LC.}\label{figure4}
\end{figure}

\begin{figure}
\begin{center}
\includegraphics[width=18cm,height=10cm]{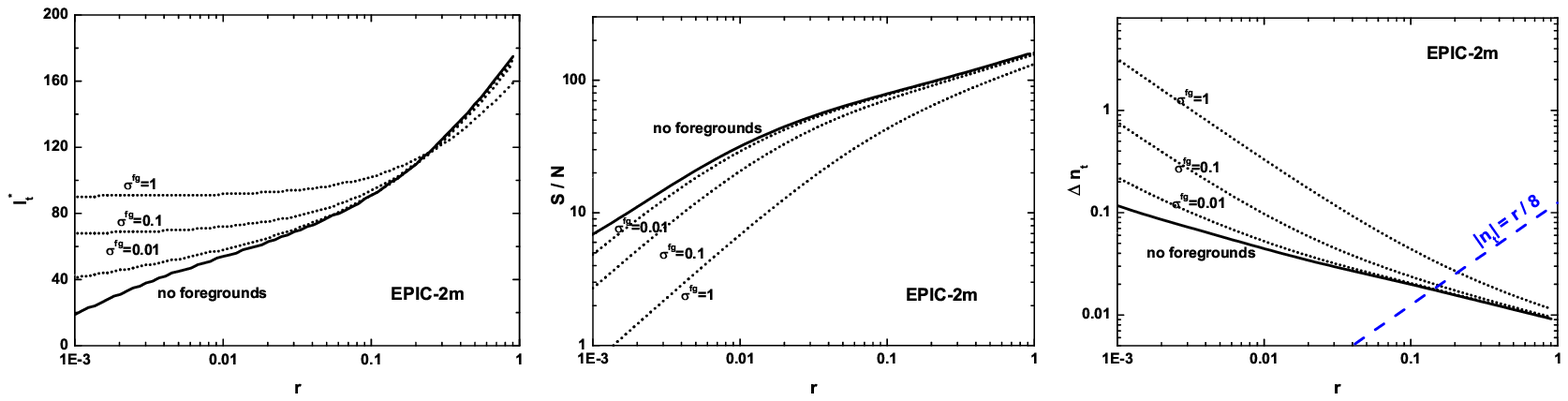}
\end{center}\caption{ This figure shows the values of $\ell_t^*$, $S/N$ and $\Delta n_t$
depend on the foreground contaminations for
EPIC-2m.}\label{figure5}
\end{figure}

\begin{figure}
\begin{center}
\includegraphics[width=18cm,height=10cm]{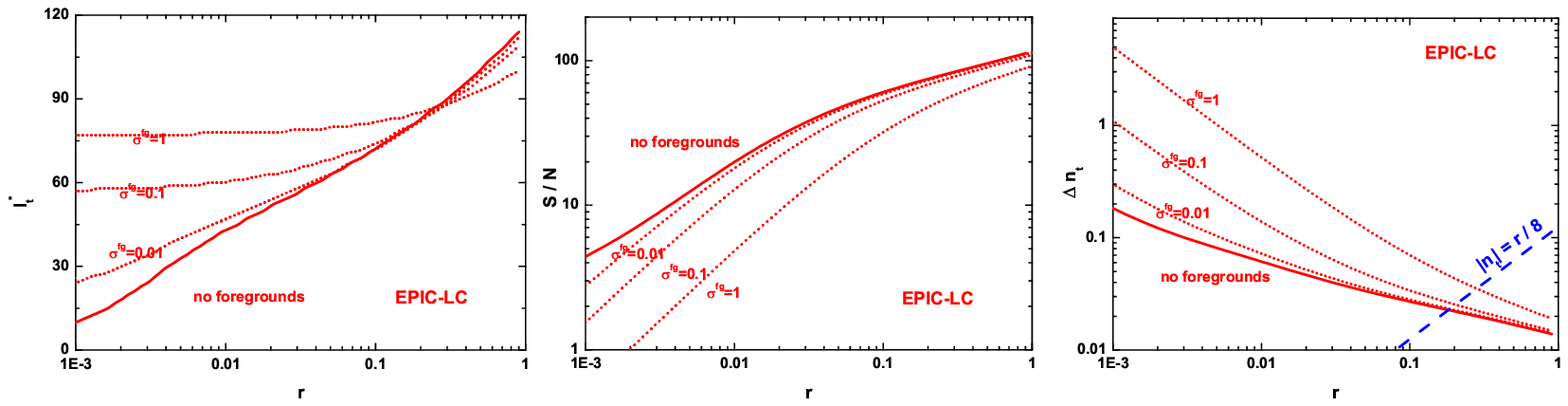}
\end{center}\caption{ This figure shows the values of $\ell_t^*$, $S/N$ and $\Delta n_t$
depend on the foreground contaminations for
EPIC-LC.}\label{figure6}
\end{figure}

\begin{figure}
\begin{center}
\includegraphics[width=16cm,height=10cm]{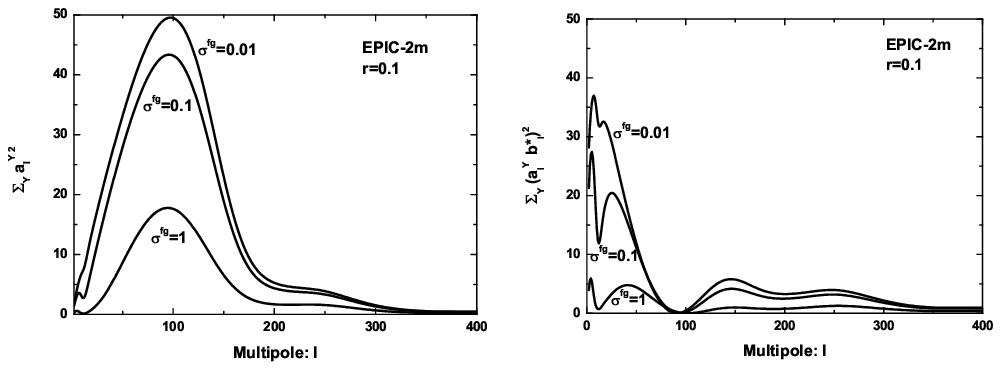}
\end{center}\caption{The figures show the values of $\sum_{Y}a_{\ell}^{Y2}$ (left panel) and
$\sum_{Y}(a_{\ell}^{Y}b^*_{\ell})^2$ (right panel) as functions of
multipole $\ell$ for the cases with different $\sigma^{\rm
fg}$.}\label{figure7}
\end{figure}

In this section, besides the instrumental noises and cosmic
lensing contaminations, we shall take into account the impact of
polarized foregrounds on the future CMBPol mission. In this paper,
we shall neglect the effect of foregrounds on the CMB temperature,
as the foreground cleaning is expected to leave a negligible
contribution in the temperature \cite{bennett}. CMB polarized
foregrounds arise due to free-free, synchrotron, and dust
emission, as well as due to extra-galactic sources such as radio
sources and dusty galaxies. In this paper, we shall only consider
only synchrotron and dust emission, which are expected to be
dominant in the CMBPol frequency range \cite{cmbpol-fore}.

The synchrotron emission results from the acceleration of
cosmic-ray electrons in the magnetic field of Galaxy, which has
been well measured on large angular scale at 23 GHz by WMAP.
Following \cite{cmbpol,verde,cmbpol-fore}, for the frequency
$\nu$, the scale-dependence of the synchrotron signal may be
parameterized as
 \bea\label{syn}
 C_{{\rm
 S},\ell}^{Y}(\nu)=A_S\left(\frac{\nu}{\nu_0}\right)^{2\alpha_S}\left(\frac{\ell}{\ell_0}\right)^{\beta^Y_S}.
 \ena
The parameters in this formula for the various power spectra are
all listed in Table \ref{table2}, where $\alpha_S=-3$ is assumed,
$\beta^{E}$, $\beta^{B}$ and $\beta^{C}$ are the corresponding
$\beta^Y_S$ for synchrotron emissions. This choice matches the
synchrotron emission at $23$GHz observed and parameterized by WMAP
\cite{wmap-fore}, and agrees with the DASI measurements
\cite{dasi-fore}.

Galactic emission in the $100-6000$ GHz frequency range is
dominated by the thermal emission from warm interstellar dust
grains. Our knowledge of polarized dust emission is relatively
poor, which is expected to be characterized by the Planck
satellite in the near future. In this paper, we shall adopt the
parameterized formula for the dust emission at frequency $\nu$ as
follows, as suggested by \cite{verde,cmbpol},
 \bea\label{dust}
 C_{{\rm D},\ell}^{Y}(\nu)=p^2A_D\left(\frac{\nu}{\nu_0}\right)^{2\alpha_D}\left(\frac{\ell}{\ell_0}\right)^{\beta_D^{Y}}
 \left[\frac{e^{h\nu_0/kT}-1}{e^{h\nu/kT}-1}\right]^2,
 \ena
where $p=5\%$ and $T=18$K. We list the other parameters for the
various power spectra in Table \ref{table2}.

Various methods have been discussed to subtract the foregrounds by
their frequency-dependence (see for instance \cite{various-fore}).
In this paper, we shall not discuss the subtraction of the
foreground from the signal. Instead, as the previous works
\cite{verde,cmbpol} we assume that the foreground substraction can
be done correctly down to a given level, and treat these residual
foregrounds as a kind of known gaussian noises in the data
analysis. However, here we should mention that in this case we
have assumed one can model and subtract the power spectra of
residuals perfectly, and avoid the possible issues of bias. This
might be a huge challenge for the future polarization observation.

If we consider the CMB experiment, including several frequency
channels, and the different channels have different noise levels,
the optimal channel combination gives the effective noise power
spectra \cite{verde,cmbpol}
 \bea\label{foreground}
 [N_{{\rm eff},\ell}^{Y}]^{-1}=\sum_{i}\left[N_{{\rm
 fg},\ell}^{Y}(i)+N_{{\rm ins},\ell}^{Y}(i)\right]^{-1},
 \ena
where $i$ runs though the channels. $N_{{\rm ins},\ell}^{Y}(i)$ is
the instrumental noise power spectra of channel $\nu_i$. $N_{{\rm
fg},\ell}^{Y}(i)$ is the residual foreground noises of channel
$\nu_i$, which is
 \bea\label{fore}
 N_{{\rm fg},\ell}^{Y}(i)=\sum_{{\rm fore=S,D}}C_{{\rm fore},\ell}^{Y}(\nu_i)\sigma^{{\rm
 fg}}+{\cal N}_{{\rm fg},\ell}^{Y}(\nu_i).
 \ena
Here, $C_{{\rm fore},\ell}^{Y}$ is the model for the power
spectrum of the synchrotron and dust signals at the frequency
$\nu_i$ given by Eqs. (\ref{syn}) and (\ref{dust}), and
$\sigma^{{\rm fg}}$ is the assumed residual factor. ${\cal
N}_{{\rm fg},\ell}^{Y}(\nu_i)$ is the noise power spectrum of the
foreground template map, as foreground templates are created by
effectively taking map differences and thus are somewhat affected
by the instrumental noise. This term can be calculated by
\cite{verde,cmbpol}
 \bea
 {\cal N}_{{\rm fg},\ell}^{Y}(\nu_i)=\frac{{N'}_{{\rm ins},\ell}^{Y}(\nu_{\rm ref})}{n_{\rm chan}(n_{\rm
 chan}-1)/4}\left(\frac{\nu_i}{\nu_{\rm ref}}\right)^{2\alpha},
 \ena
where $n_{\rm chan}$ is the total number of channels used, and the
reference channel $\nu_{\rm ref}$ is the highest and lowest
frequency channel included in the cosmological analysis for dust
and synchrotron respectively, i.e., that listed in Table
\ref{table1}. The parameters $\alpha$ for the foregrounds under
consideration are defined in Table \ref{table2}, i.e. $\alpha=-3$
for the synchrotron emissions and $\alpha=2.2$ for the dust
emission. The quantity ${N'}_{{\rm ins},\ell}^{Y}(\nu_{\rm ref})$
is the white instrumental noise (without the beam window function)
of the corresponding template channel \cite{verde}.

Thus the total noise power spectra, by combining the
multipole-frequency instrumental noises and the residual
foregrounds, as well as the residual cosmic lensing contamination,
are given by
 \bea\label{total}
 N_{\ell}^{X} \mapsto N_{{\rm
 eff},\ell}^{X}~(X=T,C,E),~~N_{\ell}^{B} \mapsto N_{{\rm
 eff},\ell}^{B}+C_{{\rm lens},\ell}^{B}\times \sigma^{\rm lens},
 \ena
where $\sigma^{\rm lens}$ is the residual factor for cosmic
lensing contamination. In this and the following sections, we
adopt $\sigma^{\rm lens}=0.5$ for EPIC-2m, and $\sigma^{\rm
lens}=1$ for EPIC-LC. The effective noise power spectra $N_{{\rm
eff},\ell}^{Y}$ strongly depend on the residual factor
$\sigma^{\rm fg}$ for the foregrounds. When no foreground
subtraction is assumed, we have $\sigma^{\rm fg}=1$. In this
paper, we also consider two assumed residual cases, suggested by
CMBPol team \cite{cmbpol}: $\sigma^{\rm fg}=0.01$ for the
optimistic case, and $\sigma^{\rm fg}=0.1$ for the pessimistic
case. In Fig. \ref{figure4}, we plot the effective noise power
spectrum $N_{{\rm eff},\ell}^{B}$ with different $\sigma^{\rm fg}$
for EPIC-2m (left panel) and EPIC-LC (right panel). We find that
for both EPIC-2m and EPIC-LC, the foregrounds increase the
effective noise power spectrum in all the multipole range when
$\sigma^{\rm fg}=1$. However, when the foregrounds can be well
subtracted, the residual foregrounds only increase the noise in
the large scale. For $\sigma^{\rm fg}=0.1$, the effective noise is
increased in the range $\ell<300$, and for $\sigma^{\rm fg}=0.01$,
the effective noise is increased in the range $\ell<100$. We find
that even if the optimistic case with $\sigma^{\rm fg}=0.01$ is
realized, the effective noise is much larger in the reionization
peak ($\ell<20$) comparing with the no foreground case.
Especially, when $r$ is small, this increased noise is larger than
the signal $C_{\ell}^B$, and decreases the contribution of the
reionizaiton peak. We have emphasized above, the reionization peak
is very important for the constraint of spectral index $n_t$ for
the CMBPol mission, it is predictable that the value of $\Delta
n_t$ would become much larger due to the foreground
contaminations, even if the optimistic case is considered. This
will be clearly shown in the following discussion.

By using the total effective noise power spectra in (\ref{total}),
we can calculate the values the best-pivot multipole $\ell_t^*$,
the signal-to-noise ratio $S/N$, and the uncertainty of the
spectral index $\Delta n_t$. In Fig. \ref{figure5} and Fig.
\ref{figure6}, we show the results for EPIC-2m and EPIC-LC,
respectively. We find that so long as $r<0.3$, the value of
$\ell_t^*$ is increased with the increasing of the residual
foregrounds. This is because that, the contaminations from
foreground are mainly in the low multipole (see Fig.
\ref{figure4}). Increasing the foregrounds, the contribution for
the detection of RGWs in the low multipoles becomes less and less
important, induces an increasing of $\ell_t^*$.

From Fig. \ref{figure5} and Fig. \ref{figure6}, we find that when
$\sigma^{\rm fg}=0.01$, the optimistic case we considered, the
foregrounds decrease the $S/N$ and increase $\Delta n_t$, when the
tensor-to-scalar ratio $r<0.03$. However, when $r>0.03$, the
effect of this residual foregrounds is negligible. This is also
easily understood. The residual foregrounds with $\sigma^{\rm
fg}=0.01$ only increase the total noise power spectra in the
largest scale $\ell<100$. This increased total noises are beyond
the signals when $r$ is small. We also find that, when
$\sigma^{\rm fg}=0.1$ or $\sigma^{\rm fg}=1$, i.e. the foregrounds
are not well subtracted, the effect of the foreground
contaminations are quite important, especially for the
determination of $n_t$. With the decreasing of $r$, the
contaminations become more and more important. For the EPIC-2m and
the input model with $r=0.1$, when $\sigma^{\rm fg}=0.01$, we have
$\Delta n_t=0.020$. When $\sigma^{\rm fg}=0.1$, the value becomes
$\Delta n_t=0.024$, and when $\sigma^{\rm fg}=1$, the value
becomes $\Delta n_t=0.045$, two times larger than that in the
optimistic case. We can investigate another case, for the EPIC-2m
and the input model with $r=0.01$, when $\sigma^{\rm fg}=0.01$, we
have $\Delta n_t=0.054$. When $\sigma^{\rm fg}=0.1$, the value
becomes $\Delta n_t=0.098$, and when $\sigma^{\rm fg}=1$, the
value becomes $\Delta n_t=0.330$, six times larger than that in
the optimistic case. So we conclude that if the value of $r$ is
not too small, such as $r=0.1$, we do not need to remove the
foreground to a very high level. The difference between optimistic
case and pessimistic is very small. However, if the value of $r$
is smaller than $0.01$, very detailed removal for the foregrounds
is very important for the determination of RGWs.

Fig. \ref{figure5} and Fig. \ref{figure6} also show that in the
optimistic case, EPIC-2m can detect the signal of RGWs with
$r=0.001$ at 5-$\sigma$ level, and EPIC-LC can detect it at
3-$\sigma$ level. However, in the pessimistic case, EPIC-2m can
only detect this signal at 2.8-$\sigma$ level, and EPIC-LC can
detect it at 1.6-$\sigma$ level.

As in the previous sections, we can discuss the contribution to
the $S/N$ and $\Delta n_t$ from the individual multipole, by
investigating the functions $\sum_{Y}a_{\ell}^{Y2}$ and
$\sum_{Y}(a_{\ell}^{Y}b_{\ell}^*)^2$. They are plotted in Fig.
\ref{figure7}, where we have considered the EPIC-2m, and the model
with $r=0.1$. We find that the foreground contamination mainly
affects the $S/N$ by decreasing the value of
$\sum_{Y}a_{\ell}^{Y2}$ around the peak at $\ell\sim100$. However,
it affects the value of $\Delta n_t$ mainly by decreasing of
$\sum_{Y}(a_{\ell}^{Y}b_{\ell}^*)^2$ at the largest scale
$\ell<50$ and the intermedial range $\ell\sim200$.

Now, let us investigate the possible application of the CMBPol
mission to differentiate the different inflationary models, which
plays a role for the future inflation researches. As well known,
one of the most important ways to distinguish different classes of
inflations is to test the so-called inflationary consistency
relations. This testing strongly depends on the determination of
the parameters specifying the relic gravitational waves, i.e. the
tensor-to-scalar ratio $r$ and the spectral index $n_t$. Now, let
us focus on the possible testing of the consistency relation for
the canonical single-field slow-roll inflationary models
\cite{liddle}. This testing might provide the unique
model-independent criteria to confirm or rule out this class of
models. The possible testing for other inflationary models by the
CMBPol mission and the ideal CMB experiment can be found in the
recent work \cite{zhao-huang}.

The consistency relation for the canonical single-field slow-roll
inflationary models can be written as \cite{liddle}
 \bea\label{consistency}
 r=-8 n_t.
 \ena
We find that this relation only depends on the parameters $r$ and
$n_t$. Since the absolute value of $n_t$ is expected to be one
order smaller than that of $r$, and also the measurement of $n_t$
is much more difficult than $r$, how well we can measure the
spectral index $n_t$ plays a crucial role for testing the
consistency relation in Eq. (\ref{consistency}).

To access whether CMBPol mission might achieve the consistency
relation test goal, in Fig. \ref{figure5} and Fig. \ref{figure6}
(right panels), we compare the values of $|n_t|=r/8$ with $\Delta
n_t$. If $\Delta n_t<|n_t|$, then the constraint on $n_t$ is tight
enough to allow for the testing. From Fig. \ref{figure5}, we find
that for the EPIC-2m mission, $\Delta n_t<|n_t|$ is satisfied only
if $r>0.14$ for the optimistic case with $\sigma^{\rm fg}=0.01$.
In the pessimistic case with $\sigma^{\rm fg}=0.1$, it becomes
$r>0.15$. Similar results for the EPIC-LC can be found in Fig.
\ref{figure6}. $\Delta n_t<|n_t|$ is satisfied only if $r>0.18$
for the optimistic case, and $r>0.20$ for the pessimistic case. So
we conclude that the testing of the consistency relation for the
canonical single-field slow-roll by the CMBPol mission is quite
hard. The testing is possible only for some large-field
inflationary models. However, we should mention that the situation
could become quite promising for the general Lorentz-invariant
single-field inflations and the two-field inflations (see
\cite{zhao-huang} for the details).

\section{Systematics contaminations \label{section6}}

Beyond raw sensitivity requirements for the instrumentals, and the
removal of the astrophysical foregrounds, much attention has
already been given in the literature to the instrumental
systematics for the constraints of the cosmological parameters and
the cosmic weak lensing reconstruction
\cite{sys1,sys3,keating,dea}. The main goal of this section is to
illustrate the effect of the instrumental systematics and
systematically study the impact on the gravitational waves
detection for the CMBPol mission.

All the effects of the beam systematics are associated with beam
imperfections or beam mismatch in dual beam experiments. Several
of these effects (e.g. differential gain, differential beam width
and the first order pointing error) are reducible with an ideal
scanning strategy and otherwise can be cleaned from the data set.
Other spurious polarization signals, such as those due to
differential ellipticity of the beam, second order pointing errors
and the differential rotation, persist even in the case of ideal
scanning strategy and perfectly mimic CMB polarization.

The beam systematics due to optical imperfections are dependent of
the underlying sky, the properties of the polarimeter and the
scanning strategy. If the outputs of two beams with orthogonal
polarization-sensitive directions are slightly different, the
temperature anisotropy can leak to the polarization or the
$E$-mode polarization can leak to the $B$-mode and vice verse.
(see \cite{keating} for the details). For example, if two beams
are exactly same but the overall response, this difference of the
measured intensity can generate a non-vanishing polarization
signal. Another typical example is effect of the beam rotation,
which is caused by the uncertainty in the overall beam
orientation. This effect mixes the Stokes parameters $Q$ and $U$,
and induces a $E$-mode and $B$-mode leakage.

The CMB power spectra $C_{{\rm sys},\ell}^{Y}$ generated by these
systematics are discussed in details by a number of authors. In
the work \cite{keating}, the authors discussed these effects
separately, and got the simple analytical formulas to calculate
the leading order of the generated power spectra, which are listed
in Table \ref{table3}. The formulae in Table \ref{table3}
separately describe the effects of the following instrumental
systematics for experiments with the elliptical gaussian beams:
differential gain effect, monopole effect, differential pointing
effect, quadrupole effect, differential rotation effect.
Differential gain can induce spurious polarization singles from
temperature leakage due to beam mismatch. This effect is described
by the parameter $g\equiv g_1-g_2$, where $g_1$ and $g_2$ refer to
the gain factors of first and second beams. The differential
rotation effect is due to uncertainty in the overall beam
orientation. This mixes the $Q$ and $U$ Stokes parameters and as a
result leaks $E$ to $B$ and vice verse. We describe this effect by
the parameter $\varepsilon \equiv
(\varepsilon_1+\varepsilon_2)/2$, where $\varepsilon_1$ and
$\varepsilon_2$ are rotation errors of first and second beams. We
note that these two parameters $g$ and $\varepsilon$ are not
related to the beam shape. The monopole effect arises from
circular beams with unmatched main-beam full width at half
maximum, which is described by the parameter
$\mu\equiv(\sigma_1-\sigma_2)/(\sigma_1+\sigma_2)$, where
$\sigma_1$ and $\sigma_2$ are the mean beamwidthes of first and
second beams. The quadrupole effects arises from beams with
differential ellipticities, and described by the ellipticity
parameter ${\rm e}\equiv (\sigma_x-\sigma_y)/(\sigma_x+\sigma_y)$,
where $\sigma_x$ and $\sigma_y$ are the major and minor axes of
the beam. Also, differential pointing, i.e. the dipole effect, is
described the parameter $\rho\equiv \rho_1-\rho_2$, where $\rho_1$
and $\rho_2$ are the circular positions of first and second beams.
These three effects can induce future spurious polarization
singles from temperature leakage.

In the formulae in Table \ref{table3}, the functions $f_1$, $f_2$
and $f_3$ are experiment-specific and encapsulate the information
about the scanning strategy which couples to the beam mismatch
parameters to generate spurious polarization. The exact
definitions of $f_1$, $f_2$ and $f_3$ are given in  Eq. (27) in
\cite{keating}, i.e.
 \[
 f_1=\frac{1}{2}|\tilde{h}_+(-1,0)|^2,~~f_2=\frac{1}{2}|\tilde{h}_+(-1,-1)|^2+\frac{1}{2}|\tilde{h}_+(-1,1)|^2,
 ~~f_3=\frac{1}{2}\langle\tilde{f}(0,1)\tilde{h}^*_-(1,-1)\rangle,
 \]
where
 \[
 f(m,n)\equiv \langle
 e^{-i(2m+n)\alpha}\rangle,~~h_{\pm}\equiv\frac{1}{D}[f(m,n)-f(m\pm2,n)\langle
 e^{\pm4i\alpha}\rangle],~~D\equiv1-\langle e^{4i\alpha}\rangle
 \langle
 e^{-4i\alpha}\rangle.
 \]
$\tilde{f}(m,n)$ and $\tilde{h}_{\pm}(m,n)$ are the Fourier
transformations of $f(m,n)$ and $h_{\pm}(m,n)$. Exact definition
of $\alpha$ can be found in \cite{keating}. Angular brackets
represent average over measurements of a single pixel, averaged
over time. In general, these functions are spatially-anisotropic
but for simplicity, and to obtain a first-order approximation, we
consider them constants in general.

In the real data analysis, these generated CMB power spectra may
be treated as a part of the signals, as well as the real signals
generated by the perturbations fields, i.e.
 \bea
 C_{\ell}^Y\mapsto C_{\ell}^Y+C_{{\rm sys},\ell}^Y.
 \ena
Thus the estimator of this contamination $D_{\ell}^{Y}$ becomes
biased for the true power spectra $C_{\ell}^Y$, by a term $C_{{\rm
sys},\ell}^Y$. And this biased estimator will induce the biased
estimator for the cosmological parameters in the likelihood
analysis. As in general, we can use the $r^*_{\rm ML}$ and
$n_{t{\rm ML}}$ as the best estimator for the parameters $r^*$ and
$n_t$. Following the previous works \cite{dea}, we define the bias
of the tensor-to-scalar ratio $\delta r$ and the spectral index
$\delta n_t$ as follows,
 \bea
 \delta r\equiv\langle r^*_{\rm
ML}\rangle-r^*,~~\delta n_t\equiv\langle n_{t{\rm ML}}\rangle-n_t.
 \ena
Given the beam systematics the bias of the tensor-to-scalar ratio
$\delta r$ and the spectral index $\delta n_t$ can be calculated
by the following formulae (similar to the previous works
\cite{dea}),
 \bea\label{deltas}
 \delta r=
r^*\frac{\sum_{\ell}\sum_{Y}a_{\ell}^Ye_{\ell}^Y}{\sum_{\ell}\sum_{Y}a_{\ell}^{Y2}},~~
 \delta n_t=
\frac{\sum_{\ell}\sum_{Y}a_{\ell}^Ye_{\ell}^Yb_{\ell}^*}{\sum_{\ell}\sum_{Y}(a_{\ell}^{Y}b_{\ell}^*)^2},
 \ena
where $e_{\ell}^Y\equiv C_{\ell}^Y({\rm
sys})/\sigma_{D_{\ell}^Y}$. Notice that, for the requirement of
the beam systematics of CMBPol mission (for instant, the parameter
$g$, associated with the differential gains, satisfies
$g\ll0.01\%$ \cite{cmbpol-lensing}), the power spectra generated
by the beam systematics are expected to be much smaller than those
generated by gravitational waves with $r>10^{-3}$ but the very
large multipole range, where the noises are dominant. So beam
systematics cannot change the value of the best-pivot multipole.

Considering the CMB experiment with multi-frequency channels, the
effective combined noise power spectra in Eq. (\ref{foreground})
can be extended to the follows, considering the contribution of
the systematics,
 \bea\label{systematics}
 [N_{{\rm eff},\ell}^{Y}]^{-1}=\sum_{i}\left[N_{{\rm
 fg},\ell}^{Y}(i)+N_{{\rm ins},\ell}^{Y}(i)+C_{{\rm sys},\ell}^{Y}(i)
 \right]^{-1},
 \ena
where $i$ runs though the channels. Throughout this section, we
shall assume the optimistic foreground removal with the residual
factor $\sigma^{\rm fg}=0.01$ for both EPIC-2m and EPIC-LC
missions. We should remember that, similar to the previous
discussion, the $B$-mode contamination ($\sigma^{\rm lens}=0.5$
for EPIC-2m, and $\sigma^{\rm lens}=1$ for EPIC-LC) by weak
lensing effect is also considered throughout this section. The
total CMB power spectra generated beam systematics can be
calculated by $C_{\ell}^{Y}({\rm sys})=N_{{\rm eff},\ell}^{Y}({\rm
with~sys})-N_{{\rm eff},\ell}^{Y}({\rm no~sys})$. In the simplest
case with single frequency channel, this term returns to
$C_{\ell}^{Y}({\rm sys})=C_{{\rm sys},\ell}^Y$.

\begin{figure}
\begin{center}
\includegraphics[width=8cm,height=8cm]{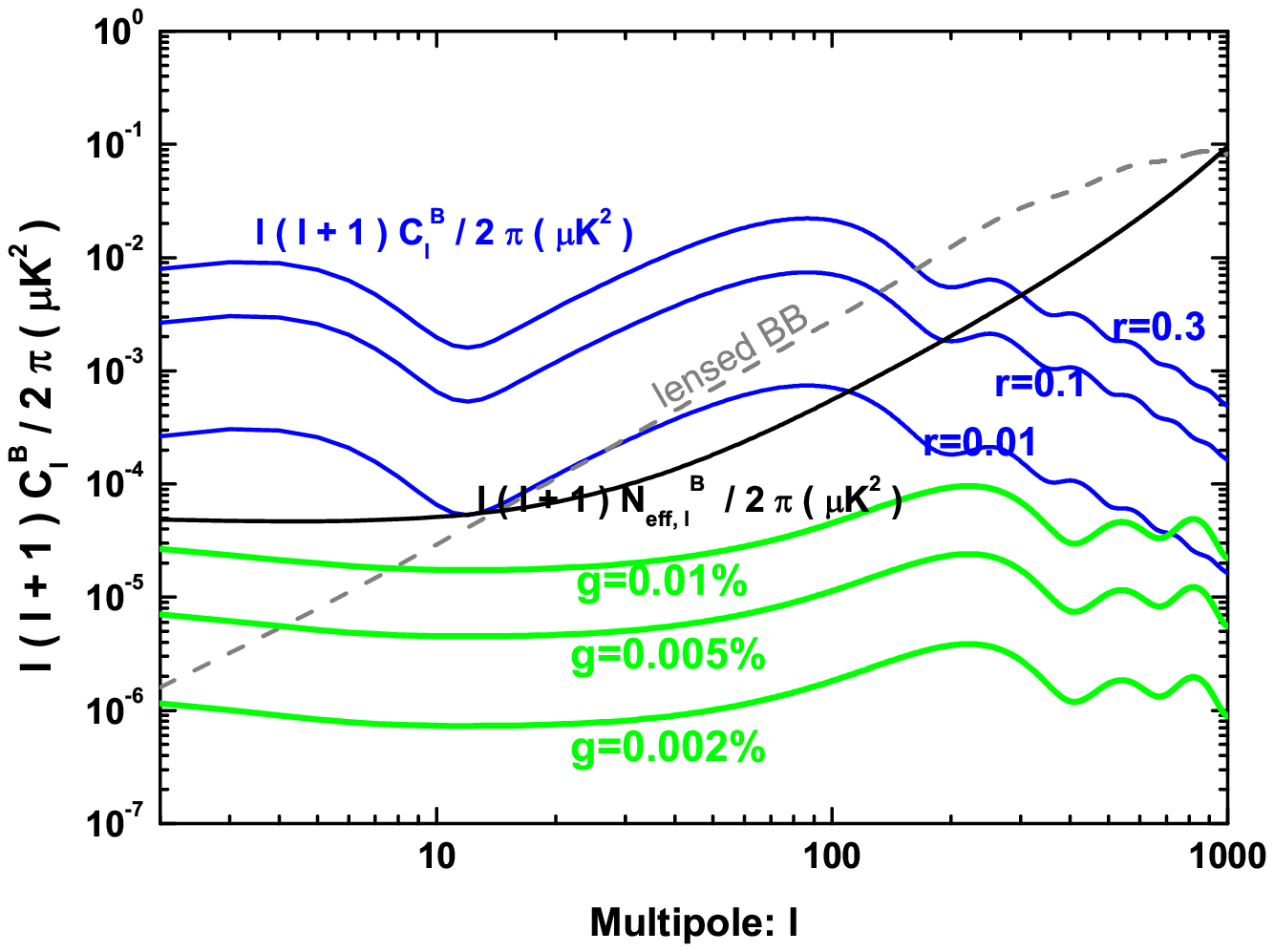}\includegraphics[width=8cm,height=8cm]{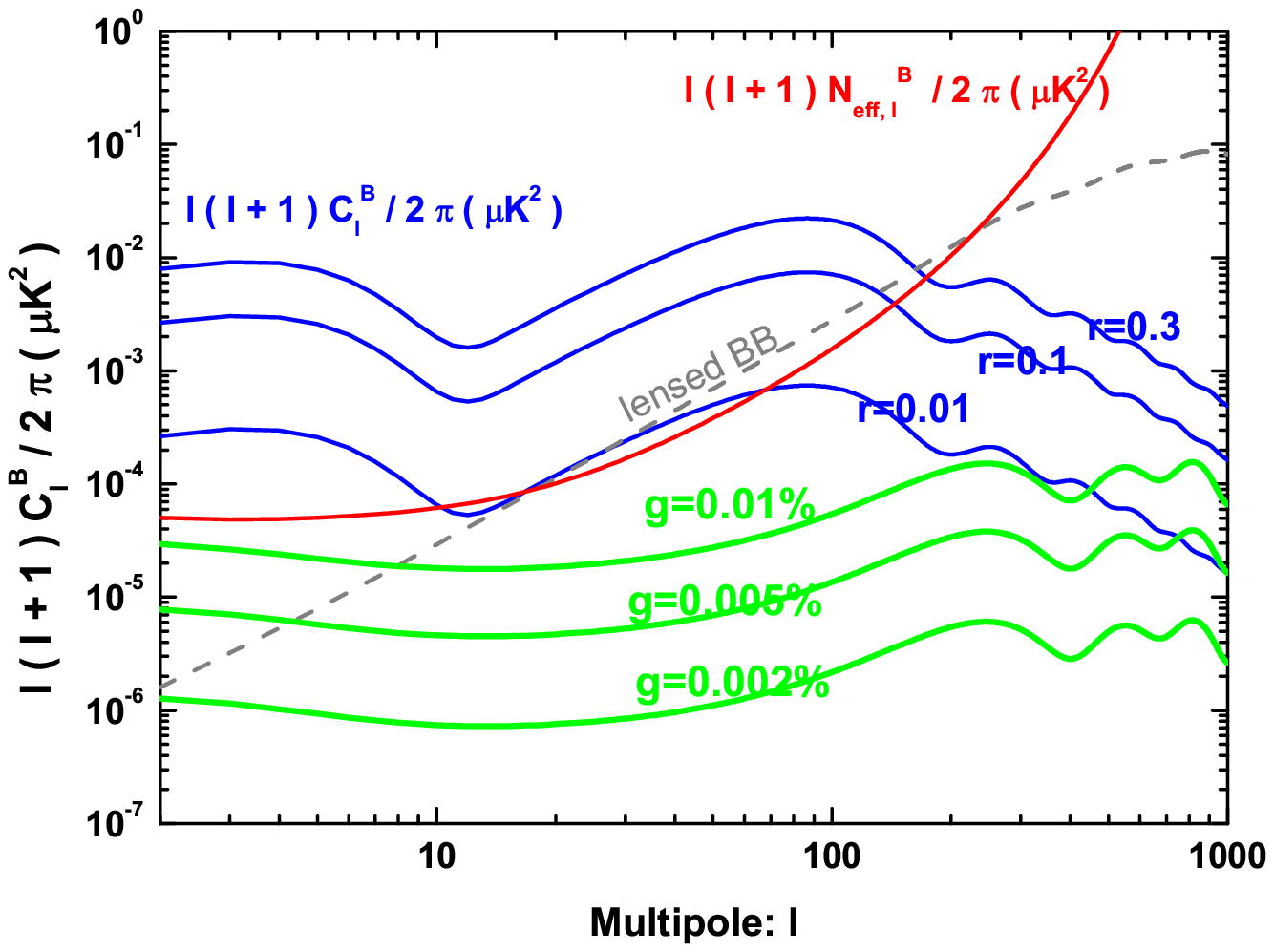}
\end{center}\caption{ The contribution of differential gain to the $B$-polarization,
comparing with the signal of $C_{\ell}^{B}$ for different $r$, the
noise of CMBPol mission, and the lensed $B$-polarization. Left
panel is for EPIC-2m, and the right panel is for
EPIC-LC.}\label{figure8}
\end{figure}

\begin{figure}
\begin{center}
\includegraphics[width=16cm,height=8cm]{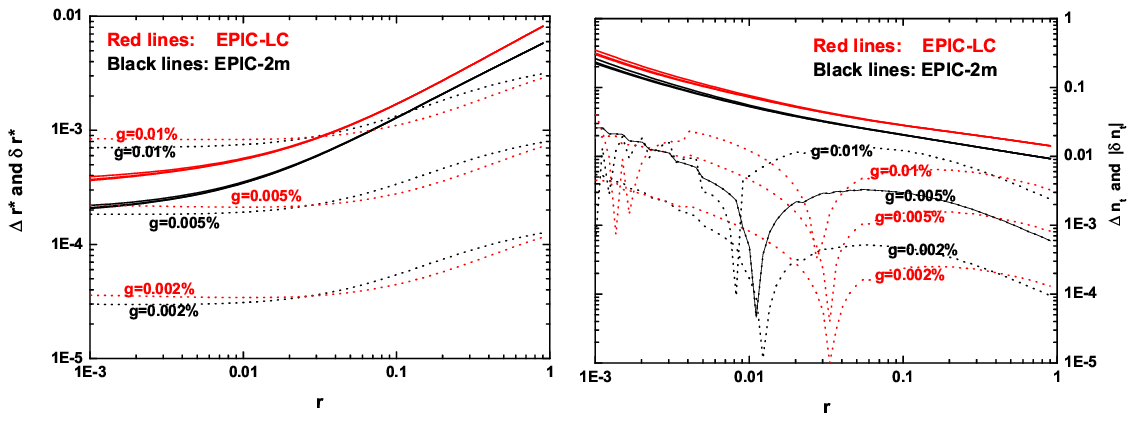}
\end{center}\caption{The values of ${\delta}r$ (left panel) and $\delta n_t$ (right panel) from the differential
gain for different tensor-to-scalar ratio $r$. For the comparison,
we plot the corresponding $\Delta r$ and $\Delta n_t$ in solid lines in the panels.}\label{figure9}
\end{figure}

Let us separately investigate the five systematical effects. Fig.
\ref{figure8} shows $C_{\ell}^{B}({\rm sys})$ for different values
of the parameter $g=0.002\%$, $0.005\%$ and $0.01\%$. In all these
figures, $f_1=2\pi$ is used as the worst case. The values of
$\ell(\ell+1)C_{\ell}^{B}({\rm sys})$ only weekly depends on the
multipole $\ell$. As long as $r>0.01$, we find $C_{\ell}^{B}({\rm
sys})$ are all smaller than those of the signals $C_{\ell}^{B}$ or
the noises $N_{{\rm eff},\ell}^B$ in the range $\ell<200$. In Fig.
\ref{figure9}, we plot the values of biases $\delta r$ and $\delta
n_t$ induced by the differential gains. We find that the values of
$\delta r$ and $\delta n_t$ strongly depends on the value of the
parameter $g$. A larger $g$ follows a larger bias. However, the
uncertainties $\Delta r$ and $\Delta n_t$ are nearly independent
of $g$ unless the value of $g$ is too large. We also find that,
given the parameter $g$ the value of the ratio $\delta r/\Delta r$
also strongly depend on the tensor-to-scalar ratio $r$. For
example, the EPIC-2m experiment with $g=0.01\%$, $\delta r$ is
larger than $\Delta r$ when $r<0.04$, the bias is very obvious.
However when $r>0.04$, we have $\delta r<\Delta r$, the bias is
smaller than the uncertainty.

Similar to the previous work \cite{dea}, we can define the
critical value $g_c$, which is the largest value of $g$ as long as
the the condition $\delta r/\Delta r<0.1$ is satisfied. In Table
\ref{table4}, we list the critical values of $g_c$ for EPIC-2m and
EPIC-LC, where $r=0.001$, $0.01$ and $r=0.1$ are considered. Since
Fig. \ref{figure9} shows that for a given $g$ value, both $\Delta
r$ and $\delta r$ increase with the increasing of $r$. So the
tendency of the critical $g_c$ for different input $r$ is not
trivial. After analysis, we find that for the assumed $g$ value,
with the increasing of $r$, if the increasing of $\Delta r$ is
more rapid than that of $\delta r$, thus the case with larger $r$
corresponds to a larger $g_c$. This is clearly shown in Table
\ref{table4}. On the other hand, if the increasing of $\delta r$
is more rapid than that of $\Delta r$, thus the case with larger
$r$ corresponds to a smaller $g_c$. Similar discussion is also
applied to the $n_t$ case, as well as the cases for the other four
systematics contaminations. From Table \ref{table4}, we find the
severest constraints are obtained from the requirement of the
small $r$ case. The requirement for EPIC-2m is quite close to that
of EPIC-LC.

Let us discuss the effect of beam gain on the determination of
spectral index $n_t$. From Eq. (\ref{deltas}), we find that the
bias $e_{\ell}^Y$ in the lower multipole range $\ell<\ell_t^*$
contributes a negative $\delta n_t$, and the bias $e_{\ell}^Y$ in
the high multipole range $\ell>\ell_t^*$ contributes a positive
$\delta n_t$. These two components are cancelled by each other and
total bias $\delta n_t$ is expected to be very small, which is
clearly shown in Fig. \ref{figure9} (right panel). Comparing with
the bias of $r$, the bias $\delta n_t$ is much smaller than that
of $\Delta n_t$. So the constraint on $g$ obtained from the
requirement of the $n_t$ is much larger than that from the
parameter $r$.

\begin{figure}
\begin{center}
\includegraphics[width=8cm,height=8cm]{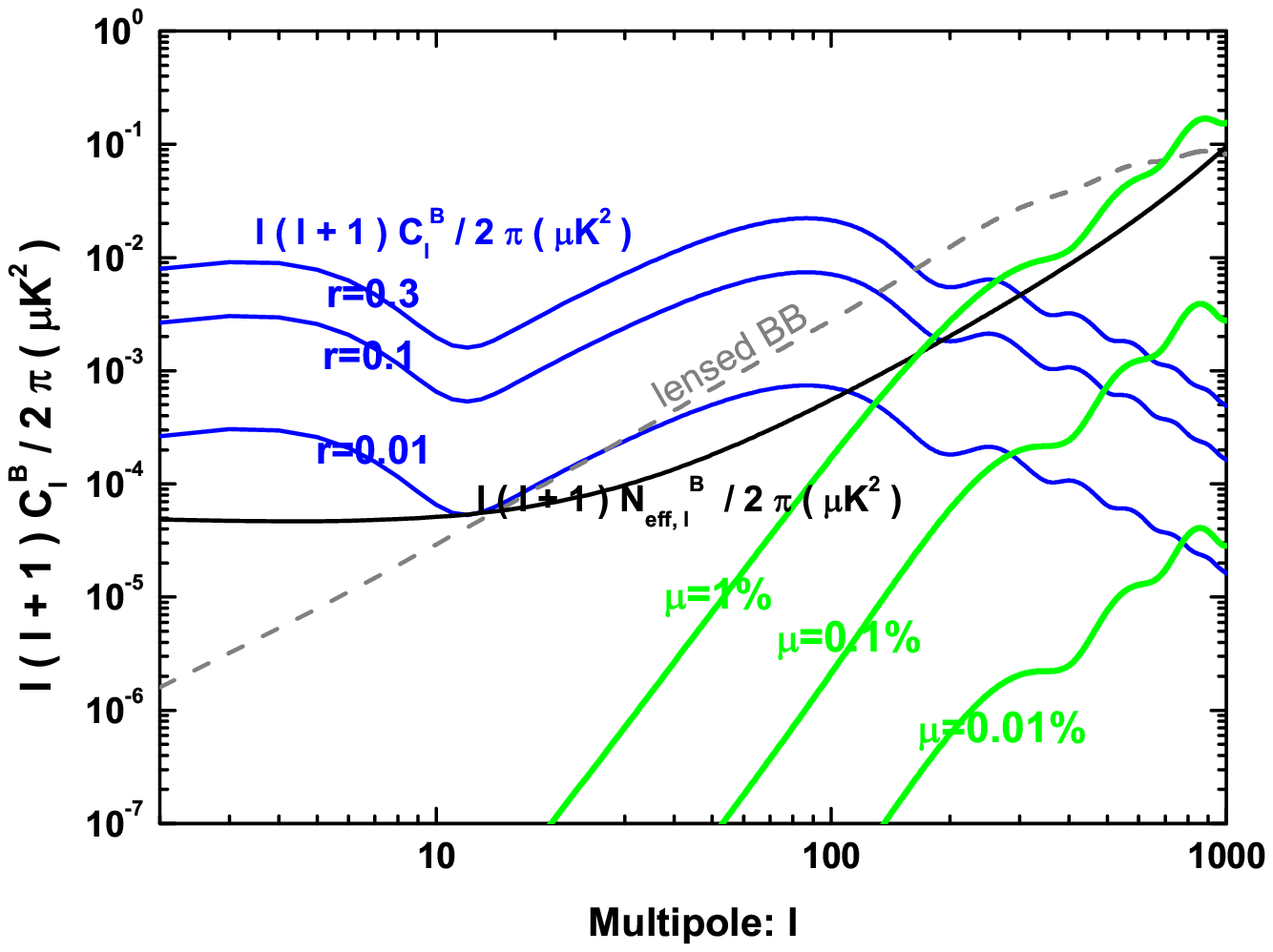}\includegraphics[width=8cm,height=8cm]{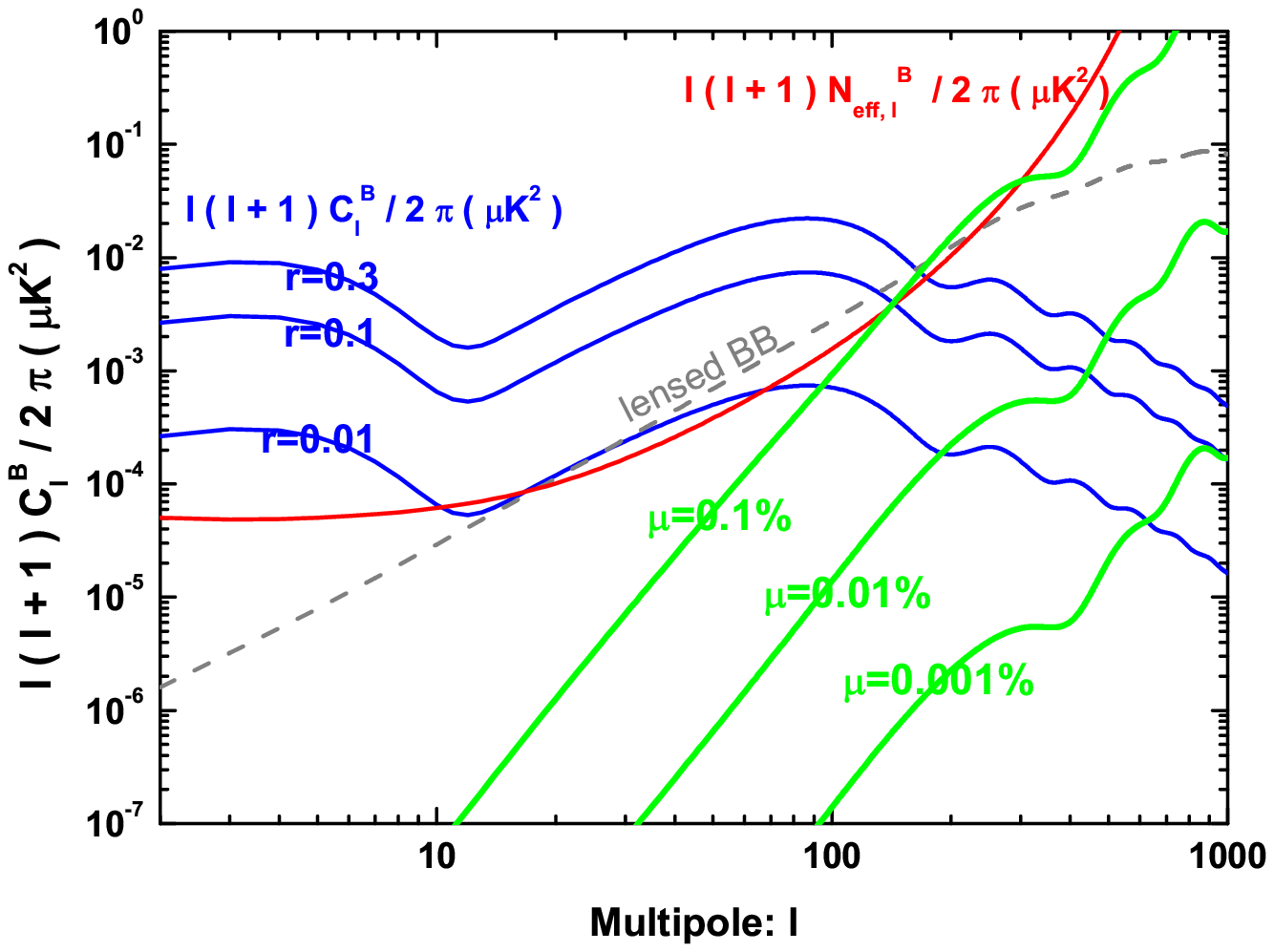}
\end{center}\caption{The contribution of monopole effect to the $B$-polarization, comparing with the signal
of $C_{\ell}^{B}$ for different $r$, the noise of CMBPol mission,
and the lensed $B$-polarization. Left panel is for EPIC-2m, and
the right panel is for EPIC-LC.}\label{figure10}
\end{figure}

\begin{figure}
\begin{center}
\includegraphics[width=16cm,height=8cm]{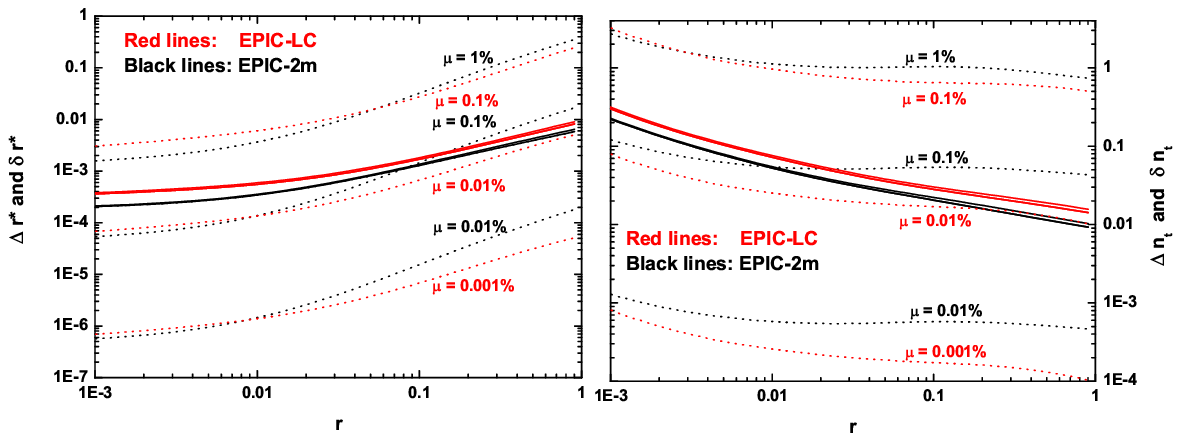}
\end{center}\caption{The values of ${\delta}r$ (left panel) and $\delta n_t$ (right panel) from the
monopole effect for different tensor-to-scalar ratio $r$. For the
comparison, we plot the corresponding $\Delta r$ and $\Delta n_t$
in solid lines in the panels.}\label{figure11}
\end{figure}

Now, let us turn to the effect of the differential monopole
effect. From the formulae in Table \ref{table3} we know the power
spectra generated by differential monopole effect strongly depend
on the beam size. Larger beam size follows the larger power
spectra $C_{{\rm sys},\ell}^Y$. This can be seen clearly in Fig.
\ref{figure10}. Given $\mu=0.1\%$, the value of $C_{\ell}^{B}({\rm
sys})$ is much larger in EPIC-LC than that in EPIC-2m. So in order
to achieve a same value of $\delta r/\Delta r=0.1$, the
requirement for EPIC-LC is much severer than that for EPIC-2m,
which are clearly shown in Table \ref{table4}. Fig. \ref{figure11}
shows that, for a given $\mu$, a larger $r$ follows a larger ratio
value $\delta r/\Delta r$, which is correct for both EPIC-2m and
EPIC-LC. In order to keep the tensor-to-scalar ratio in the range
$r\in(0.001,0.1)$ unbiased, we should have $\mu_c=0.029\%$ for
EPIC-2m, and $\mu_c=0.005\%$ for EPIC-LC, where $f_1=2\pi$ is
adopted. we can also discuss the effect of differential monopole
on the determination of spectral index $n_t$. From Fig.
\ref{figure10}, we find that $C_{{\rm sys},\ell}^Y$ is sharply
peaked at the high multipole, where $\ell>\ell_t^*$. Thus the
total contribution to $\delta n_t$ is always positive (see Fig.
\ref{figure11}), especially when the value of $\mu$ is not too
small, the effect of differential monopole at the small scale is
very important, and follows a fairly large bias for spectral
index. Let us define the value of $\mu_c$, where $\delta
n_t/\Delta n_t=0.1$ is satisfied. From Table \ref{table4} we find
that the constraint on $\mu$ is a little severer from the
parameter $n_t$ than that from the parameter $r$. This table shows
that, the most severe constraint on $\mu$ is obtained from the
requirement of $n_t$ in the case of $r=0.1$.

\begin{figure}
\begin{center}
\includegraphics[width=8cm,height=8cm]{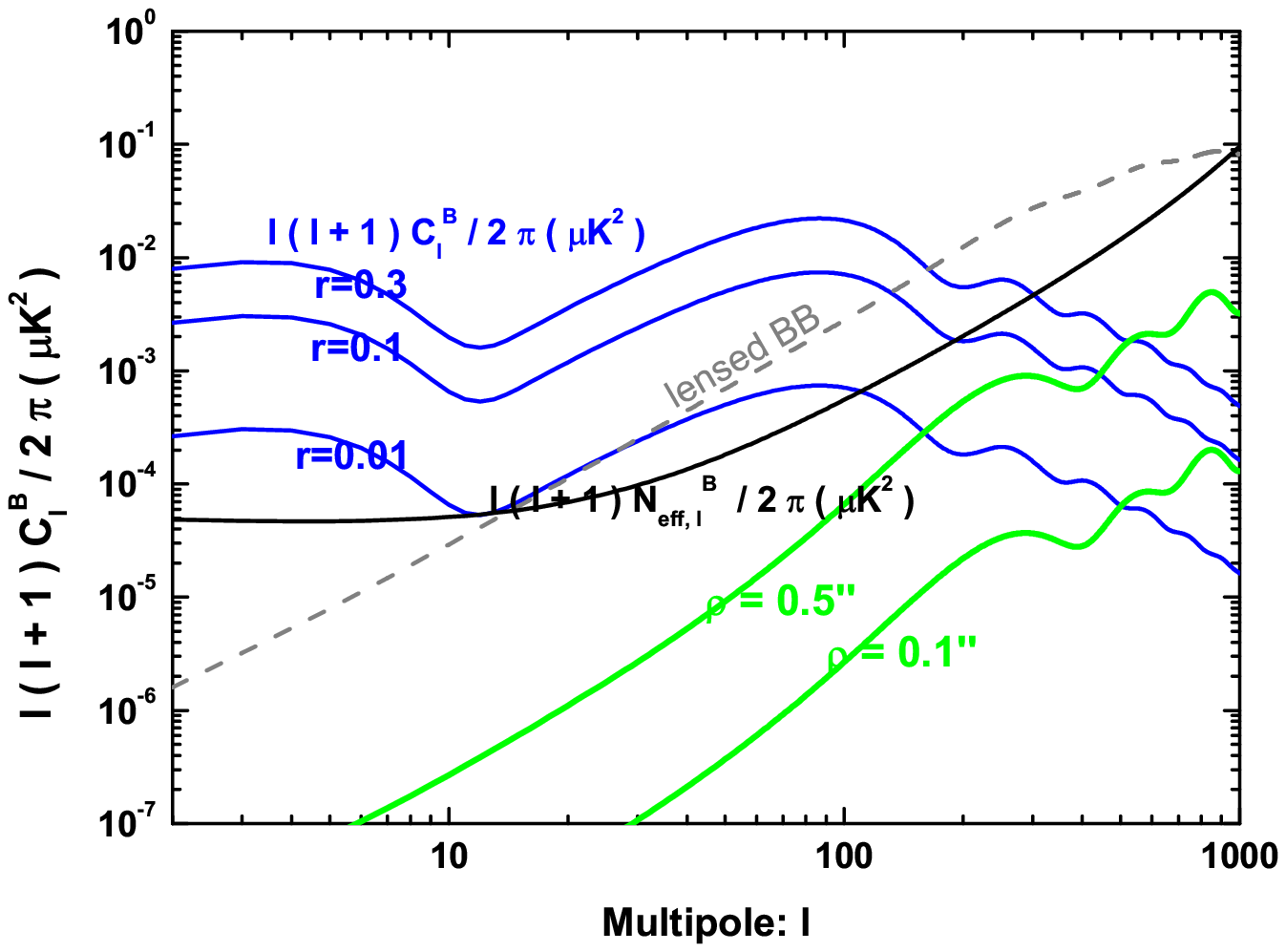}\includegraphics[width=8cm,height=8cm]{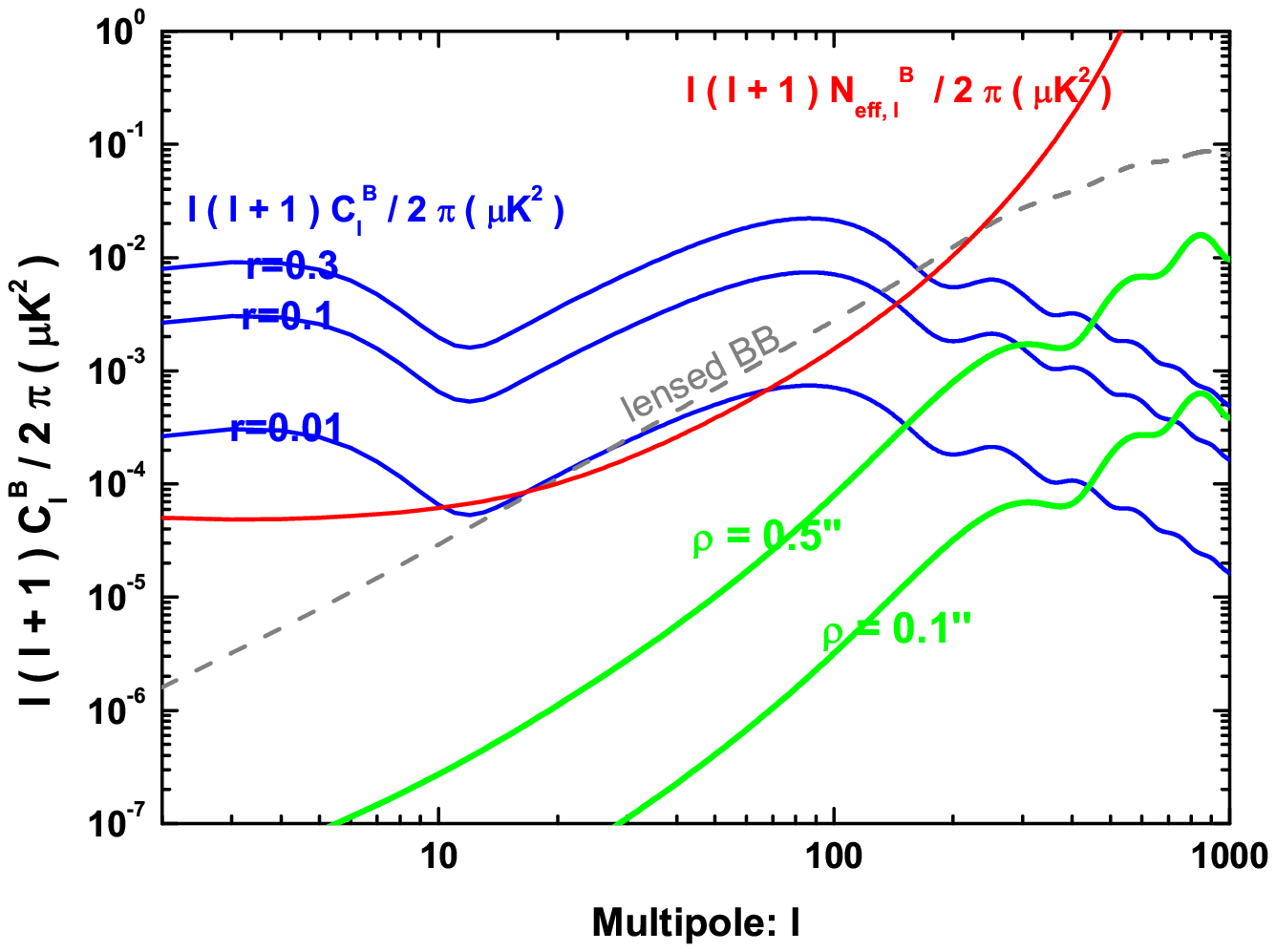}
\end{center}\caption{The contribution of differential pointing to the $B$-polarization,
comparing with the signal of $C_{\ell}^{B}$ for different $r$, the
noise of CMBPol mission, and the lensed $B$-polarization. Left
panel is for EPIC-2m, and the right panel is for
EPIC-LC.}\label{figure12}
\end{figure}

\begin{figure}
\begin{center}
\includegraphics[width=16cm,height=8cm]{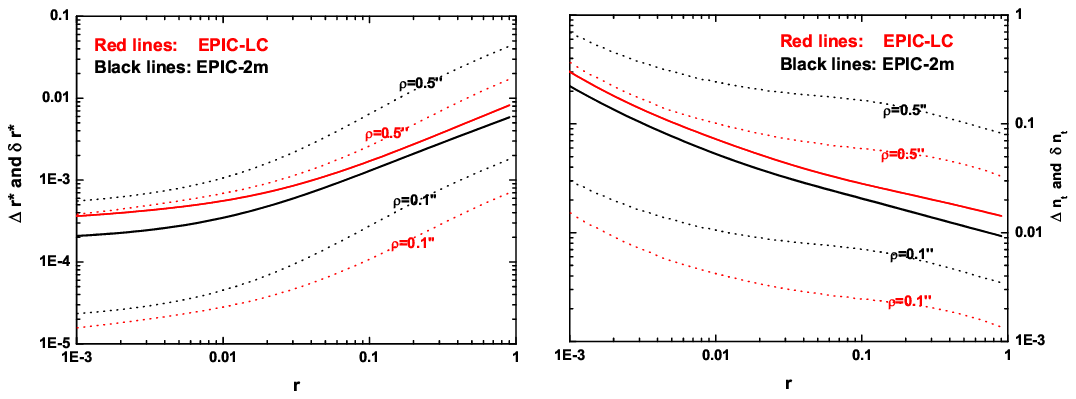}
\end{center}\caption{The values of ${\delta}r$ (left panel) and $\delta n_t$ (right panel) from the
differential pointing for different tensor-to-scalar ratio $r$. For the comparison,
we plot the corresponding $\Delta r$ and $\Delta n_t$ in solid lines in the panels. }\label{figure13}
\end{figure}

In Figs. \ref{figure12} and \ref{figure13}, we show the effect of
differential pointing on the determination of gravitational waves,
where $f_2=2\pi$ is adopted. These figures show that similar with
the case of differential monopole effect, the function
$C_{\ell}^{B}({\rm sys})$ generated by differential pointing is
also sharply peaked at the high $\ell$, which follows that the
bias of the spectral index $\delta n_t$ is positive. Comparing the
left panel with the right panel in Fig. \ref{figure12}, we find
that for a given $\rho$, the effect of the differential pointing
is more important for the EPIC-2m, due to the smaller instrumental
noises. So more severe constraint on the differential pointing is
followed for the EPIC-2m. In Table \ref{table4}, we find that for
both EPIC-2m and EPIC-LC, the most severe on $\rho$ comes from the
requirement of parameter $n_t$ at $r=0.1$.

\begin{figure}
\begin{center}
\includegraphics[width=8cm,height=8cm]{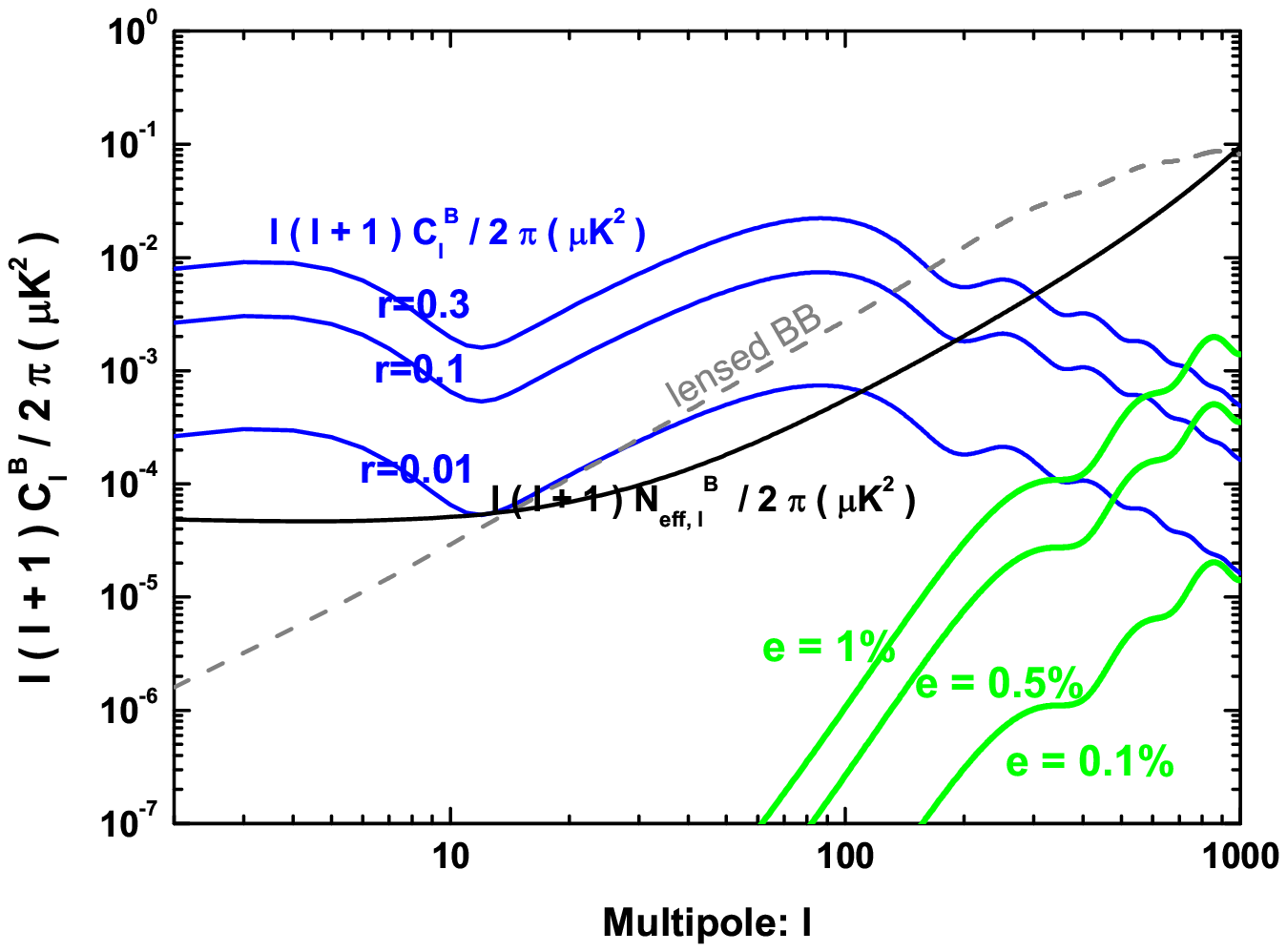}\includegraphics[width=8cm,height=8cm]{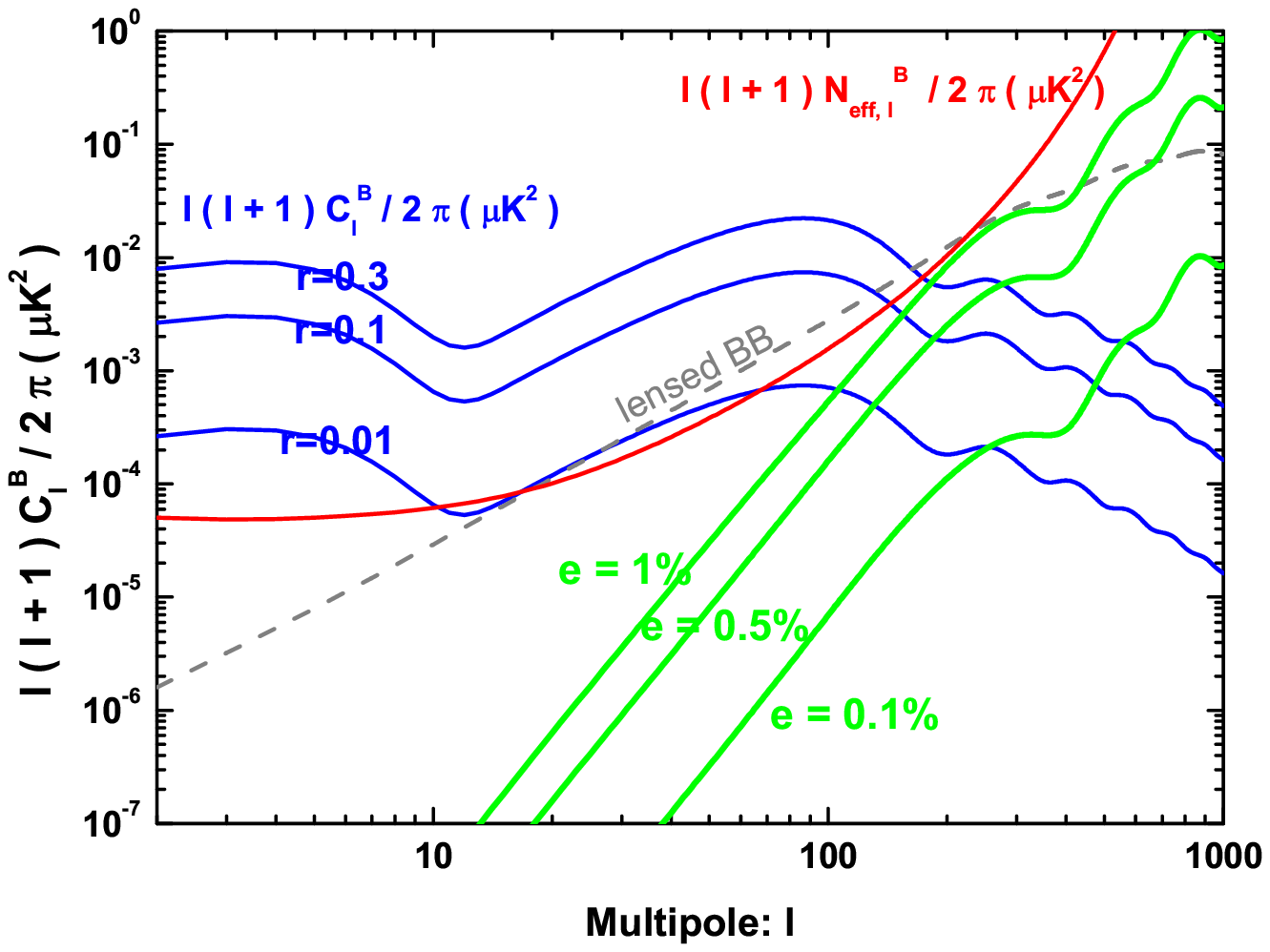}
\end{center}\caption{The contribution of quadrupole effect to the $B$-polarization,
comparing with the signal of $C_{\ell}^{B}$ for different $r$, the
noise of CMBPol mission, and the lensed $B$-polarization. Left
panel is for EPIC-2m, and the right panel is for
EPIC-LC.}\label{figure14}
\end{figure}

\begin{figure}
\begin{center}
\includegraphics[width=16cm,height=8cm]{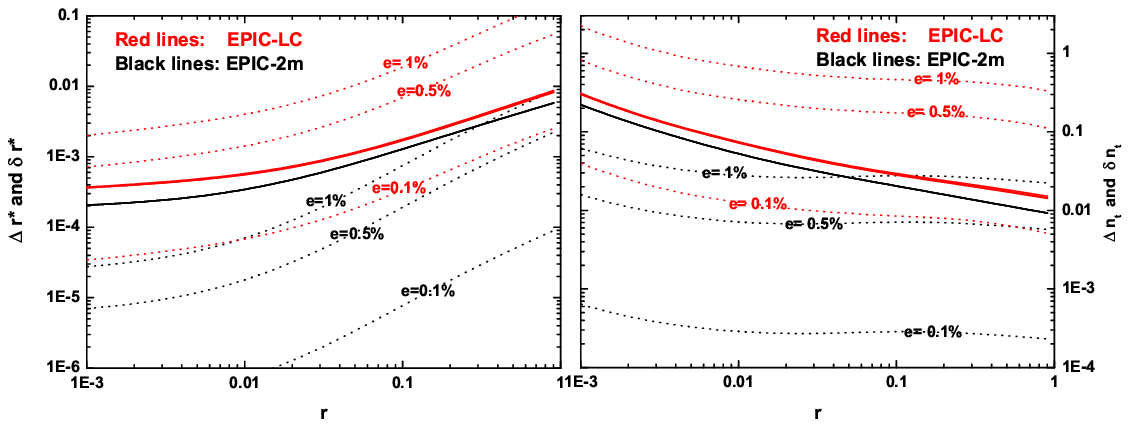}
\end{center}\caption{The values of ${\delta}r$ (left panel) and $\delta n_t$ (right panel) from
the quadrupole effect for different tensor-to-scalar ratio $r$.
For the comparison, we plot the corresponding $\Delta r$ and
$\Delta n_t$ in solid lines in the panels.}\label{figure15}
\end{figure}

We have also investigated the effect of the differential
quadrupole in Figs. \ref{figure14} and \ref{figure15}. The effects
are similar to those of the differential monopole. We find that
EPIC-LC needs the much stricter requirement than EPIC-2m. For each
experiment, the most severe constraint on the parameter ${\rm e}$
is obtained from the requirement of $n_t$ in the case with
$r=0.1$.

\begin{figure}
\begin{center}
\includegraphics[width=8cm,height=8cm]{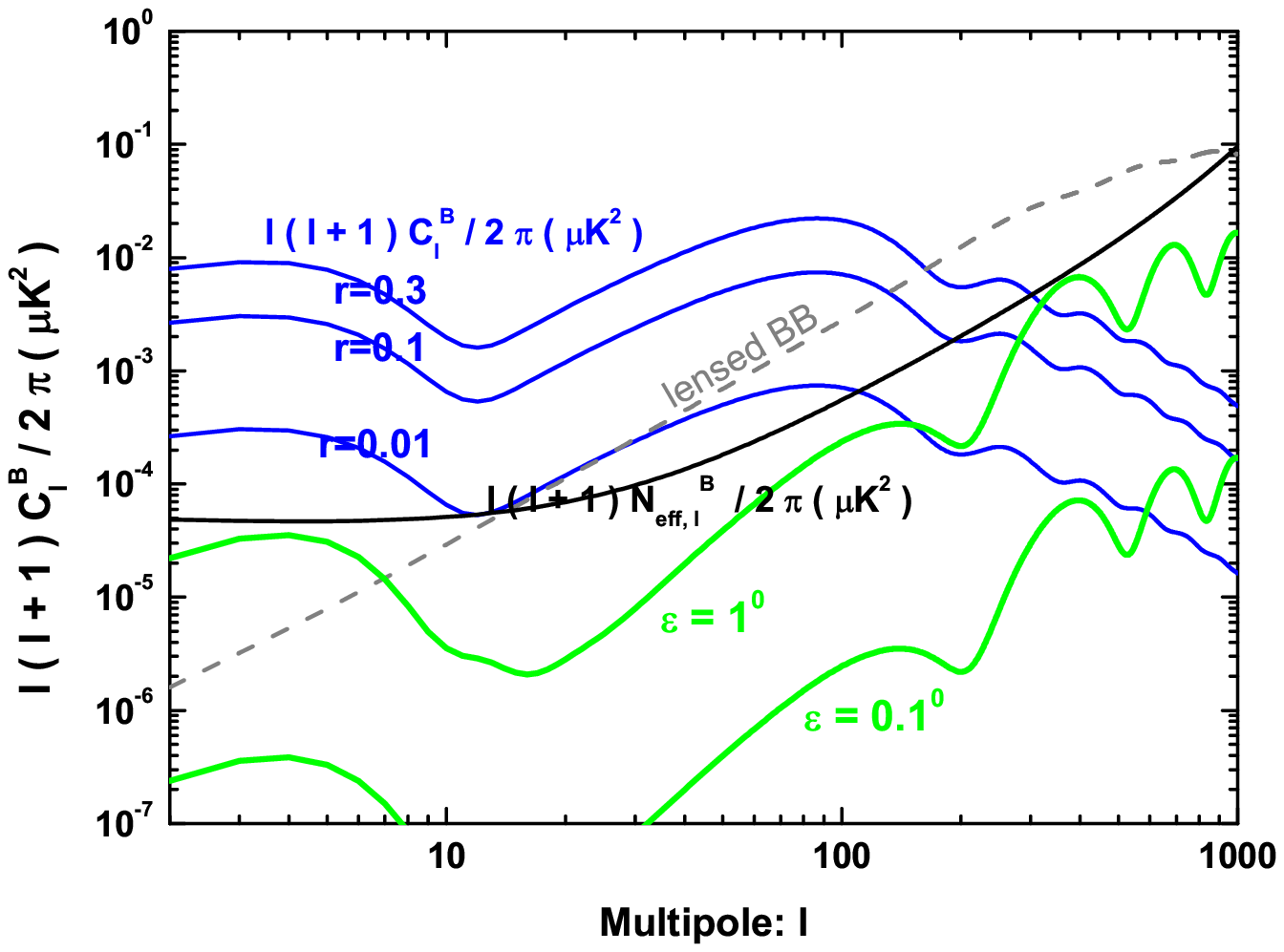}\includegraphics[width=8cm,height=8cm]{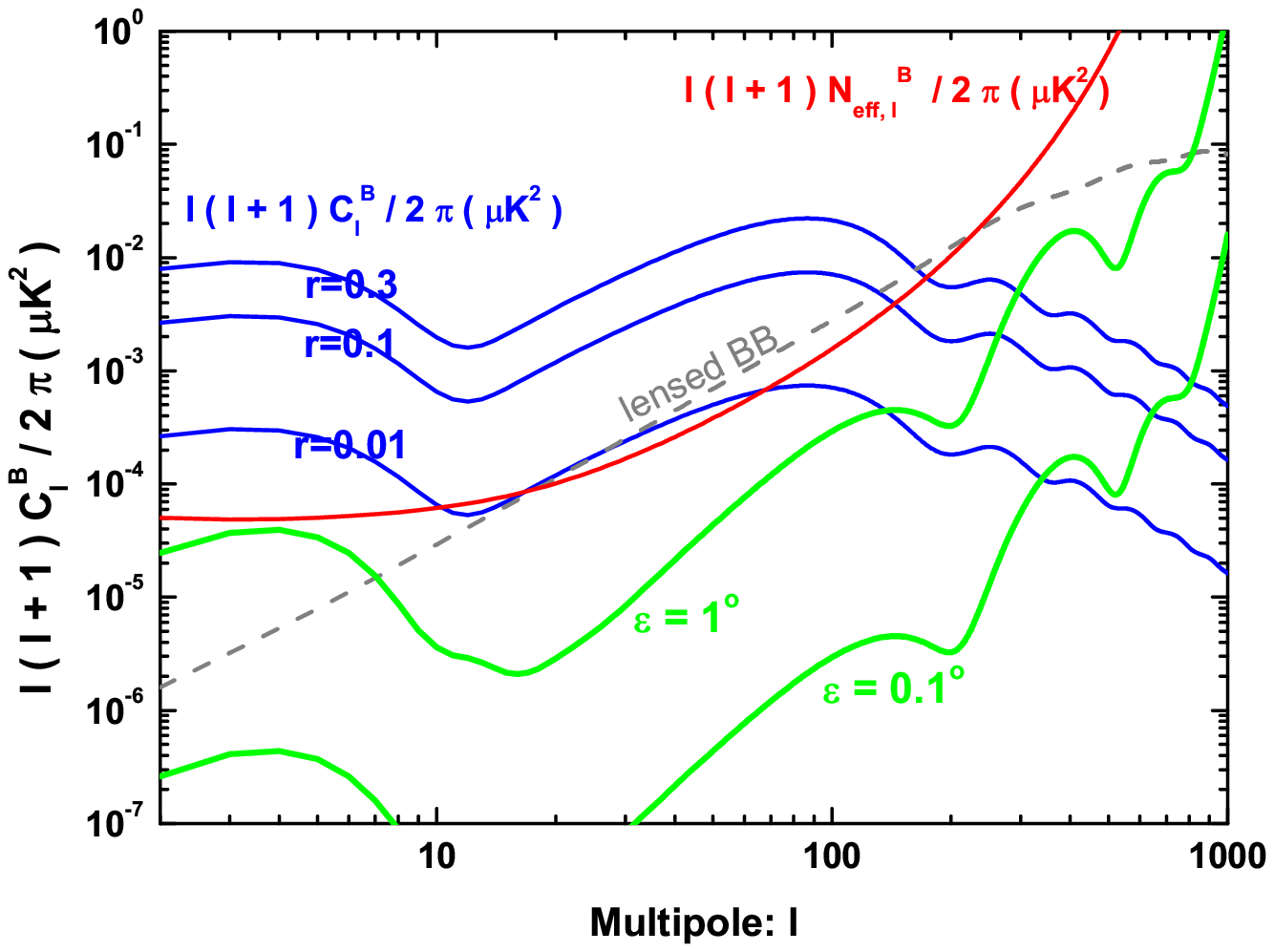}
\end{center}\caption{ The contribution of differential rotation to the $B$-polarization,
comparing with the signal of $C_{\ell}^{B}$ for different $r$, the
noise of CMBPol mission, and the lensed $B$-polarization. Left
panel is for EPIC-2m, and the right panel is for
EPIC-LC.}\label{figure16}
\end{figure}

\begin{figure}
\begin{center}
\includegraphics[width=16cm,height=8cm]{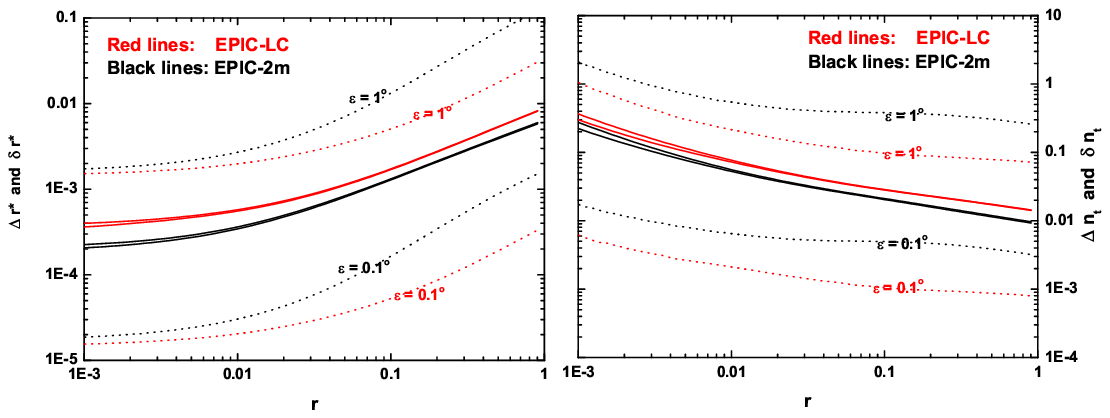}
\end{center}\caption{The values of ${\delta}r$ (left panel) and $\delta n_t$ (right panel) from the
differential rotation for different tensor-to-scalar ratio $r$. For the comparison,
we plot the corresponding $\Delta r$ and $\Delta n_t$ in solid lines in the panels.}\label{figure17}
\end{figure}

At last, we shall discuss the effect of the differential rotation,
and the results are shown in Figs. \ref{figure16} and
\ref{figure17}. We find that for a given $\varepsilon$, the ratios
$\delta r/\Delta r$ and $\delta n_t/\Delta n_t$ only weakly depend
on the tensor-to-scalar ratio. And the ratio for EPIC-2m is a
little smaller than that of EPIC-LC. In order to keep the
parameters $r$ and $n_t$ unbiased, the requirement
$\varepsilon<0.09^o$ is needed for EPIC-2m, and
$\varepsilon<0.15^o$ is needed for EPIC-LC.

As a conclusion, by analyzing the effects of the systematics on
the determination of gravitational waves, we find that the
requirement of $n_t$ unbiased follows the similar or even more
severe constraints for the beam systematical parameters. For the
effects of differential monopole, pointing and quadrupole, a
larger $r$ follows a more severe constraint for the systematics.
The critical values for the systematical parameters are listed in
Table \ref{table4}, where the bold entries denote the most severe
constraint in each case. We also find that comparing with EPIC-2m,
the low cost EPIC-LC experiment has a much high requirement for
the systematical parameters $\mu$ and ${\rm e}$.

\begin{table}
\caption{The leading order contributions of the systematic effects
to the CMB power spectra, assuming the underlying sky is not
polarized (except for the rotation signal) \cite{keating}, where
$z\equiv (\ell\sigma)^2{\rm e}$ and $\sigma$ is mean beamwidth of
the beam, which is calculated by $\sigma=\theta_F/\sqrt{8\ln 2}$.
$c_{\theta}\equiv \cos(\theta)$, where $\theta$ is the angle
between ellipse major axis of the elliptical gaussian beam and the
horizontal $x$-axis of the fixed focal plane. $c_{\psi}\equiv
\cos(\psi)$ and $s_{\psi}\equiv \sin(\psi)$, where $\psi$ is the
angle between the axis of polarization sensitivity and the major
axis of the elliptical beam. A clear show of the angles $\theta$
and $\psi$ can be found in Fig. 1 in \cite{keating}. The
definitions of the other parameters can be found in the text, see
also \cite{keating}.}
\begin{center}
\label{table3}
\begin{tabular}{|c|c|c|c|c|}
    \hline
    Effect &~~~Parameter~~~&~~~~~~~~~~$C_{{\rm sys},\ell}^{C}$~~~~~~~~~~&~~~$C_{{\rm sys},\ell}^{E}$~~~  &~~~$C_{{\rm sys},\ell}^{B}$~~~ \\
    \hline
    Gain  &$g$& 0 & $g^2f_1 C_{\ell}^{T}$  & $g^2f_1 C_{\ell}^{T}$ \\
    \hline
    Monopole   & $\mu$& 0 & $4\mu^2(\ell\sigma)^4C_{\ell}^{T} f_1$  & $4\mu^2(\ell\sigma)^4C_{\ell}^{T} f_1$ \\
    \hline
    Pointing   & $\rho$& $-c_{\theta}J_1^2(\ell\rho)C_{\ell}^{T} f_3$ & $J_1^2(\ell\rho)C_{\ell}^{T} f_2$   & $J_1^2(\ell\rho)C_{\ell}^{T} f_2$\\
    \hline
    Quadrupole   & e& $-I_0(z)I_1(z)c_{\psi}C_{\ell}^{T}$ & $I_1^2(z)c_{\psi}^2C_{\ell}^{T}$   &$I_1^2(z)s_{\psi}^2C_{\ell}^{T}$ \\
    \hline
    Rotation   &$\varepsilon$& 0 & $4\varepsilon^2 C_{\ell}^{B}$  & $4\varepsilon^2 C_{\ell}^{E}$\\
  \hline
\end{tabular}
\end{center}
\end{table}

\begin{table}
\caption{Systematics tolerance for EPIC-2m and EPIC-LC, where
different nominal values of $r$ are considered.  In each box, the
left value is for the parameter $r$ and the right value is for the
parameter $n_t$.}
\begin{center}
\label{table4}
\begin{tabular}{|c|c|c|c|c|c|c|}
    \hline
     ~&Nominal value &~~~~~$\frac{g_c}{1\%}\sqrt{\frac{f_1}{2\pi}}$~~~~~&~~~~~$\frac{\mu_c}{1\%}\sqrt{\frac{f_1}{2\pi}}
     $~~~~~&~~~~~$\frac{\rho_c}{1''}\sqrt{\frac{f_2}{2\pi}}$~~~~~&~~~~~~~$\frac{\rm e_c}{1\%}$~~~~~~&~~~~~$\varepsilon_c$[deg]~~~~~\\
    \hline
    ~&r=0.001   &{\bf 0.0017}~\&~-----&0.061~\&~0.042&0.096~\&~0.085&0.87~\&~0.59&0.11~\&~0.21\\
    EPIC-2m &r=0.01   &0.0021~\&~-----&0.049~\&~0.030&0.087~\&~0.071&0.70~\&~0.43&0.11~\&~0.14\\
    ~&r=0.1   &0.0031~\&~-----&0.029~\&~{\bf 0.019}&0.068~\&~{\bf 0.054}&0.41~\&~{\bf 0.27}&{\bf 0.09}~\&~0.10\\
    \hline
    ~&r=0.001   &{\bf 0.0020}~\&~-----&0.0073~\&~0.0061&0.15~\&~0.14&0.10~\&~0.09&{\bf 0.15}~\&~0.22\\
    EPIC-LC &r=0.01   &0.0026~\&~-----&0.0064~\&~0.0053&0.14~\&~0.13&0.09~\&~0.08&0.17~\&~0.19\\
    ~&r=0.1   &0.0039~\&~-----&0.0050~\&~{\bf 0.0040}&0.13~\&~{\bf 0.11}&0.07~\&~{\bf 0.06}&0.18~\&~0.16\\
    \hline
\end{tabular}
\end{center}
\end{table}

\section{Conclusion \label{section7}}

The proposed CMBPol mission is the next generation of the
space-based CMB experiment, which will survey the full sky and has
the much smaller instrumental noises than Planck satellite. As one
the most important tasks of this mission, detecting relic
gravitational waves will be achieved if the tensor-to-scalar ratio
$r\gtrsim0.001$, which will provide a great opportunity to study
the physics in the early Universe, especially in the inflationary
stage.

In this paper, we have detailedly discussed the detection of relic
gravitational waves by focusing on the constraints of the
parameters $r$ and the spectral index $n_t$, which are always used
to describe the primordial power spectrum of gravitational waves.
In our discussion, we deeply investigate various contaminations
for the detection, including the instrumental noises of CMBPol
mission, the cosmic lensing contamination, the foreground
contaminations, and the effect of various beam systematics. We
found that the cosmic lensing becomes the dominant noise sources,
comparing the instrumental noises for both EPIC-2m and EPIC-LC
projects. Different from Planck satellite, if $r>0.01$, the
detection of gravitational waves mostly depends on the observation
at multipole $\ell\sim100$, the peak of the $B$-mode polarization.
However, the reionization peak at $\ell\sim 10$ still plays a
crucial role for the determination of spectral index $n_t$. We
also found that if the foreground contaminations cannot be well
controlled, the reionization peak may be unobservable, which could
deeply increase the uncertainty of $n_t$. At the same time, we
have investigated the effect of various beam systematics on the
detection of gravitational waves, which mainly cause a bias on the
cosmological parameters. In order to keep these biases small
enough, the requirements for the beam systematical parameters are
quite severe, especially for the EPIC-LC mission.


\section*{Acknowledgements}

The author is partially supported by Chinese NSF Grants No.
10703005, No. 10775119, No. 11075141. We thank the anonymous
referee for the useful comments and suggestions.


.



\end{document}